\documentclass[prb,twocolumn,longbibliography,superscriptaddress,floatfix]{revtex4-2}
\usepackage{graphicx}
\usepackage{textgreek}

\usepackage{amsmath}
\usepackage{amssymb}
\usepackage{hyperref}
\usepackage[caption=false]{subfig}
\usepackage[capitalize]{cleveref}
\usepackage{braket}
\usepackage{mathtools}
\usepackage{interval}
\usepackage{bm}
\usepackage{siunitx}

\AtBeginDocument{

}
\AddToHook{cmd/appendix/before}{
    \crefalias{section}{appendix}
    \crefalias{subsection}{appendix}
}

\crefname{section}{Sec.}{Sections}

\begin{document}
\title{Superconductivity induced by spin-orbit coupling in a two-valley ferromagnet.}
\author{Zachary M. Raines}
\author{Andrey V. Chubukov}
\affiliation{School of Physics and Astronomy and William I.
    Fine Theoretical Physics Institute, University of Minnesota, Minneapolis, MN 55455, USA}
\begin{abstract}
We analyze the origin of superconductivity in a ferromagnetically ordered state of multi-layer graphene systems placed in proximity to WSe$_2$. We model these materials by a two-valley system of interacting fermions with small pockets and Ising spin-orbit coupling.
The model yields a canted ferromagnetic order, which gives rise to a half-metal.
We obtain the magnon spectrum and derive two sets of magnon-mediated 4-fermion interactions: spin-flip interactions mediated by a single magnon and spin-preserving interactions mediated by two magnons. We argue that both processes have to be included on equal footing into the magnon-mediated pairing interaction between low-energy fermions from the filled bands.
Then the full magnon-mediated interaction satisfies Adler criterion and for a valley-odd/spatially-even order parameter contains a universal attractive piece.
This term is induced by spin-orbit coupling and is confined to energies which are parametrically smaller than the Fermi energy.
We argue that, due to retardation, this magnon-mediated attraction gives rise to superconductivity despite that there exists a stronger static repulsion, in close analogy with how phonon-mediated attraction gives rise to pairing in the presence of stronger Coulomb (Hubbard) repulsion.
\end{abstract}

\maketitle

\section{Introduction}
Superconductivity in multi-layer graphene systems has attracted strong interest from the condensed matter community in the last few years, triggered by its experimental detection in twisted bilayer graphene near integer hole and electron fillings~\cite{Cao2018a,Oh2021}.
Subsequently superconductivity was also observed in
other systems, including non-twisted multi-layer graphene structures like Bernal bilayer graphene (BBG)~\cite{Zhou2022a,Zhang2023a,Holleis2023}, rhombohedral tri-layer graphene (RTG)~\cite{Zhou2021a} and, most recently,
rhombohedral penta-layer graphene (R5G)~\cite{Han2025}.
In all these systems superconductivity has been found at a finite hole/electron doping $n$ in the presence of a displacement field $D$ (an electric field, applied perpendicular to the layers).
This field splits electron-like and hole-like excitation bands and creates flat regions in momentum space at the bottom of the lowest conduction band and the top of the highest valence band.

The subject of this study is a theoretical analysis of superconductivity observed in encapsulated BBG and RTG fabricated on a WSe$_2$ substrate, which
introduces an Ising spin-orbit coupling (SOC) (Refs.~[\onlinecite{Wang2016,Island2019,Holleis2023,Zhang2023a,Patterson2024}]).
Superconductivity in these devices develops in several regions of the $(n,D)$ phase diagram and,
in at least one region, $T_c$ is much higher than in ``pure'' BBG/RTG ($\SI{300}{mK}$ vs $\SI{50}{mK}$).
Quantum oscillation measurements~\cite{Holleis2023,Zhang2023a} showed that in this ``high-$T_c$'' range, the system is a half-metal (the 4-fold spin/valley degeneracy of the Fermi surfaces is reduced to 2-fold).
Subsequent magnetometry studies~\cite{Patterson2024} showed evidence that the order is magnetic and the spin configuration is a canted
ferromagnet (CFM) --
a combination of a spontaneous FM
order of the XY spin components,
which is the same in both valleys,
and SOC-induced magnetization of $S_z$, which changes sign between the two valleys.
The highest $T_c$ develops close to the boundary of such a state, but still within it.

\begin{figure}[htb]
    \centering
    \includegraphics[width=\linewidth]{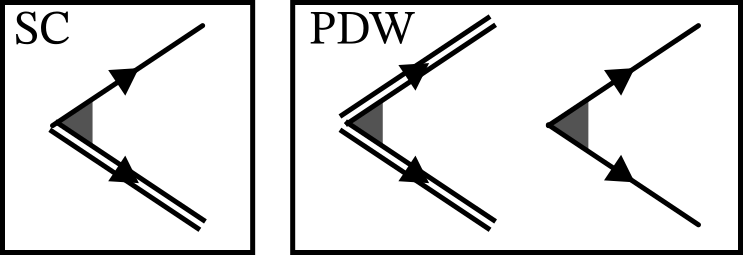}
    \caption{Types of pairing between spin-up fermions in valleys $K$ and $K'$ (solid and double solid lines).
        Left: BCS-type pairing of two fermions from different valleys with zero total momentum.
        Right: Pair density wave for fermions within the same valley, with the total momentum $2K$ or $2K'$.
        \label{fig:pairing-diagrams}}
\end{figure}

These experimental discoveries call for theoretical analysis of superconductivity coming out of a magnetically-ordered state which breaks a continuous spin symmetry ($U(1)$ symmetry in our case).
Magnetic excitations in the ordered state are different for fluctuations in the direction of order or perpendicular to it.
Longitudinal (Higgs) excitations are massive,
while transverse excitations (magnons) are massless Goldstone bosons.
In this communication we analyze whether magnons can mediate an equal-spin pairing at an elevated $T_c$.
We consider  the experimentally relevant two-valley case with low-energy fermions located near $K$ and $K'$ points in the Brillouin zone.
In this geometry, $-K = K'$ up to a reciprocal lattice vector.
Pairing of fermions within a given valley is then a pair-density-wave state with momentum $2K$ or $2K'$ (\cref{fig:pairing-diagrams}), while pairing
with zero total momentum involves fermions from both valleys.
For the latter, the Pauli principle requires that the gap function is either spatially odd and valley-symmetric (valley isospin triplet), or spatially even and valley-asymmetric (valley isospin-singlet).
We focus on isospin-singlet superconductivity at zero total momentum, which we found to be the best case scenario for magnon-mediated pairing, at least at weak/moderate coupling.

Potential superconductivity inside a magnetically ordered state, mediated by a Goldstone magnon, has been extensively analyzed for antiferromagnetic (AFM) order, in the context of cuprate superconductors (see. e.g.,
Refs.~[\onlinecite{Schrieffer1989,Schrieffer1995,Sokol1995,*Morr1997,*Morr_1997_a,Sushkov,*Sushkov_1,Ismer_2010}]).
Low-energy fermionic excitations in an AFM
remain spin-degenerate, hence a low-energy fermion, scattered by a transverse Goldstone magnon, remains near the Fermi surface.
The general argument (Adler principle) is that the
vertex involving
two
low-energy fermions and a Goldstone
boson must vanish at the Goldstone momentum $q_G$ ($(\pi,\pi)$ for a 2D AFM) to avoid fermion-induced mass generation for a Goldstone boson~\cite{Adler1965,*Watanabe2014,*Vasiliou2024}
The smallness of the vertex compensates the large value of the boson susceptibility in an AFM, and the total pairing interaction remains finite and flat for momenta near $q_G$.
\footnote{The same holds for the pairing by an acoustic phonon. The vanishing of the vertex function at $q=0$ implies that the pairing by an acoustic phonon is not advantageous compared to the one by an optical phonon.}
A similar situation holds for pairing by inter-valley Goldstone modes in magic-angle TBG~\cite{Kozii2022}
\footnote{In TBG, the low-energy fermionic excitations are nearly valley-degenerate, hence inter-valley scattering by a Golstone boson in an IVC-ordered state is again a Fermi surface process}.

The pairing interaction within a ferromagnetically (FM) ordered state
was discussed in the past~\cite{Fay1980,Kirkpatrick2001}, but recently came into focus
after the experiments on twisted and untwisted graphene monolayers and also twisted transition metal dichalcogenides~\cite{Pasupathy2024,Mak2024}.
The situation in a FM is different from that in an AFM as FM order splits spin-up and spin-down excitations.
A scattering by a FM Goldstone magnon is a spin-flip process, hence the initial and the scattered fermion cannot be simultaneously on the Fermi surface.
In this situation, a single-magnon scattering should be regarded as a non-diagonal process, and the Adler principle is not applicable.
In a recent work, \textcite{Dong2024} computed the fermion-magnon vertex function
for a single-magnon spin-flip scattering  in a CFM and found that it remains finite at the Goldstone momentum $q_G=0$.
They obtained a SOC-induced magnon-mediated interaction with two incoming spin-up fermions and two outgoing spin-down fermions,
assumed that both majority and minority carriers have Fermi surfaces (large and small Fermi pockets),
and obtained superconductivity by solving a $2\times 2$ set of gap equations on large and small pockets, in close analogy to what was done for Fe-based superconductors~\cite{Fernandes2017,*Fernandes2022}.

In this communication we propose a different pairing mechanism.
We assume that the CFM order creates a true half-metal (no minority Fermi surface) and derive the effective magnon-mediated pairing interaction involving only low-energy fermions from the majority Fermi surface, which we set to be spin-up.
Such an interaction is a combination of three processes: (i) two spin-up/spin-down scattering processes taken to second order, including the one considered in \onlinecite{Dong2024}; (ii) first-order scattering of two magnons by spin-up fermions; and (iii) the combined processes with a single two-magnon scattering and two one-magnon scatterings (\cref{fig:two-magnon-diagrams,fig:two-magnon-scattering}).
Overall, this effective interaction can be viewed as the total 2-magnon scattering process by low-energy spin-up fermions
(see~\cref{sec:effective-interaction}).

The full pairing interaction for spin-up fermions is the sum of the magnon-mediated interaction $\Gamma^{sc}_{2mag}$ and a static inter-valley density-density repulsion $U_2$, which does not contribute to CFM order and is not affected by it.
The interaction between fermions on the Fermi surface, in the limit of small momentum transfer, is
$\Gamma^{sc}_{2mag} = \int d^2d d \Omega_m A (q, \Omega_m) \chi^2 (q, \Omega_m)$, where $\chi (q, \Omega_m)$ is the magnon propagator, $q$ and $\Omega_m$ are magnon momentum and Matsubara frequency, and $A (q, \Omega_m)$ is the combined contribution from first and second-order two-magnon processes.
\footnote{One can easily make sure that processes involving more than two magnons give a smaller contribution to the pairing interaction and can therefore be neglected.}
We find that $A (0, 0)$ vanishes in line with the Adler principle~\cite{Adler1965,*Watanabe2014,*Vasiliou2024}.
We expand $A(q, \Omega_m)$ and the two magnon propagators to leading orders in $q$ and $\Omega$, integrate $A (q, \Omega_m) \chi^2 (q, \Omega_m)$ over momentum and frequency, and obtain $\Gamma^{sc}_{2mag}$ as the sum of the two terms.
One is a non-universal high-energy contribution that comes from magnons with momenta (frequencies) of order $k_F$ ($E_F$) and higher.
This contribution is essentially static and is subleading to the repulsive  $U_2$.
The other is a universal low-energy contribution that comes from
magnon momenta (frequency) parametrically smaller than $k_F$ ($E_F$).
Such a contribution is induced by the Ising SOC $\lambda$
and comes from the range of $q/\Omega_m$  parametrically smaller than $k_F/E_F$, where the magnon dispersion is linear because SOC reduces the SU(2) symmetry of a spin-isotropic FM to U(1).
We show that this universal interaction is attractive and scales parametrically with $\lambda/E_F$, which is not small in BBG and RTG proximitized to WSe$_2$ at densities where superconductivity with the highest $T_c$ has been observed~\cite{Holleis2023,Zhang2023a,Patterson2024}.
While the attractive interaction is smaller than $U_2$, we argue that it generates superconductivity due to retardation, in the same way as electron-phonon attraction generates superconductivity despite being smaller than Coulomb/Hubbard repulsion.
\footnote{A similar scenario has been proposed for superconductivity induced by either SOC or a magnetic field in the non-magnetic phase near the onset of an order~\cite{Dong2023a}.}.

\section{Model and assumptions}
\label{sec:model}
We consider a model of fermions located in the two valleys near $K$ and $K'$ points in the Brillouin zone, with Ising SOC $\lambda$ of opposite sign in the two valleys.
We set the system to be on the ``high-density'' side of the Van Hove singularity, where in the absence of order there are 4 Fermi surfaces, one per spin and per valley, and assume for definiteness an isotropic $k^2/(2m)$ fermionic dispersion.
We model the interaction between fermions by three terms: intra-valley density-density interaction $U_1$, inter-valley density-density interaction $U_2$ and inter-valley exchange interaction $U_3$.
All $U_i$ are positive (repulsive) and we assume that they already incorporate the renormalizations from fermions with energies larger than the Fermi energy $E_F$.
The interactions $U_1 \sim U_2$ describe small momentum transfer within a valley while the exchange interaction $U_3$ has momentum transfer $K-K'$.
We thus treat $U_3/U_1$ as a small parameter in our analysis.
We analyze the magnetic order, magnon propagator, magnon-fermion vertex and the effective magnon-mediated pairing interaction between low-energy fermions, all within the ladder approximation, by summing up ladder series of diagrams with particle-hole bubbles.
The computations are rather straightforward, so we skip the fine details and discuss the key intermediate steps and the results.

\section{Magnetic order and excitations}
\label{sec:mag-order}
The Ising SOC $\lambda$ acts as a valley-odd magnetic field and induces a spin polarization along the $z$ direction in spin space with opposite sign in the two valleys.
For small $U_i$, the valley staggered polarization
density
is
\begin{equation}
    \Delta_z =  \lambda \frac{N_F}{(1- (U_1-U_3)N_F)} = \frac{\lambda}{(2U_3 + (1- c)/N_F)},
\end{equation}
where we have defined $c = (U_1 + U_3)N_F$ and $N_F = m/2\pi$ is the 2D density of states.
At the critical value $c=1$, where $\Delta_z = \lambda/(2U_3)$, the system develops a spontaneous FM order in the XY spin plane; we set the order to be along $x$.
The order parameter $\Delta_x$ is the same in both valleys, hence the spin structure becomes a CFM.
\footnote{Another possibility is a spin-density-wave order with momentum $K-K'$ (a spin inter-valley coherence order).  For $U_1 + U_3 > U_2$, which we assume to hold, a homogeneous FM order is the leading instability.}.

A FM transition at $\lambda =0$ has been studied in Refs.~\onlinecite{Raines2024a,*Raines2024b,Raines2024c,*Calvera2024} and was found to be a strong first order transition from a full to half-metal: immediately past the transition the magnetization jumps to its largest possible value $\Delta_x = N_F \mu_0 = k^2_F/(4\pi)$, where $\mu_0 = E_F$ is the chemical potential before the order sets in.
Such an order completely depletes the band with the spin projection opposite to the magnetization and moves all fermions into the band with the spin projection along the magnetization.
We find that the same holds also at a finite $\lambda$:
at $c = 1+0^+$, $\Delta_x$ jumps to its maximal possible value $\Delta_x = \sqrt{(N_F \mu_0)^2 - (\lambda/2U_3)^2}$ and
the system becomes a half-metal.

Introducing $\Delta_x$ and $\Delta_z$ into the Hamiltonian, decoupling the $U_3$ term ($U_1$ and $U_2$ are not affected as they are expressed in term of full densities in a given valley, which remain intact) and diagonalizing the quadratic part, we obtain
\begin{equation}
    \mathcal{H}_2 = \sum_k E_{+} \left(f^\dag_k f_k + {\tilde f}^\dag_k {\tilde f}_k\right) +E_{-} \left(e^\dag_k e_k + {\tilde e}^\dag_k {\tilde e}_k\right)
\end{equation}
where $E_+ = k^2/(2m) -2\mu_0$, $E_- = k^2/(2m) +2\mu_0 (c-1)$, and $f ({\tilde f})$ and  $e ({\tilde e})$ are fermionic operators with spin along the magnetization and opposite to it in valley $K$ (valley $K'$).
Re-expressed in terms of the new fermions, interactions $U_1$ and $U_2$ retain their forms, while $U_3$
acquires coherence factors from the diagonalization
\begin{equation}
    \begin{split}
        \mathcal{H}_{U_3}  = U_3 \sum_{k,p,q} & \left[\sin
        \theta \left(f^\dagger_k {\tilde f}_{k+q} + e^\dagger_k {\tilde e}_{k+q}\right)\right.                                                        \\
        + {}                                  & \left. \cos {\theta} \left(f^\dagger_k {\tilde e}_{k+q} - e^\dagger_k {\tilde f}_{k+q}\right)\right]  \\
        \times {}
                                              & \left[\sin {\theta} \left({\tilde f}^\dagger_p {f}_{p-q} + {\tilde e}^\dagger_p e_{p-q}\right)\right. \\
        + {}                                  & \left. \cos {\theta}
        \left({\tilde e}^\dagger_p f_{p-q} - {\tilde f}^\dagger_p e_{p-q}\right)\right]
        \label{eq:HU3}
    \end{split}
\end{equation}
where $\theta = \arccos{\lambda/(2 U_3 N_F \mu_0)}$ is the canting angle
with respect to the $z$ axis.

CFM order spontaneously breaks U(1) symmetry in the XY plane, so for our choice of a spontaneous order along $x$,  there should be a Goldstone
mode
associated with fluctuations in the $y$ direction in spin space.
This mode has a linear dispersion, but the velocity is
non-zero only due to $\lambda$, otherwise the spin-wave dispersion is quadratic.
A convenient way to obtain the spectrum of all magnetic excitations is to introduce infinitesimal fluctuating order parameters along all three spin directions and obtain the full ones by summing up ladder series of interaction-driven renormalizations.
The ratios of the full and bare order parameters are susceptibilities, whose poles determine the dispersions of both transverse and longitudinal magnetic excitations.
In our case, there are no longitudinal excitations because the order parameter $\Delta_x$ has the maximal possible value
(the residue of the longitudinal pole vanishes).
For transverse excitations, we find after summing up ladder series that the
dispersions are the solutions of $D_{+}, D_{-} =0$, where
\begin{widetext}
\begin{equation}
    D_\pm
    = \left(1 - (U_1 \pm U_3) \Pi_A\right)  \left(1 - (U_1 \mp U_3 \cos{2\theta})\Pi_A\right)  - (U_1 \pm U_3) (U_1 \mp U_3 \cos{2\theta}) \Pi^2_B,
    \label{eq:Dplus-Dpminus}
\end{equation}
\end{widetext}
$\Pi_{A,B} = (\Pi (q, \Omega_m) \pm \Pi (q, -\Omega_m))/2$,
and $\Pi (q, \Omega_m)$ is the polarization bubble made of fermions with opposite spin projections.
Evaluating the bubbles we obtain from $D_+=0$ a U(1) gapless mode, which describes in-phase transverse fluctuations of the magnetization in $K$ and $K'$ valleys.
The mode's dispersion on the real frequency axis is
\begin{equation}
    \Omega^2 =  2\beta \mu_0 (c-1) \frac{q^2}{2m} + \left(\frac{c-1}{c}\right)^2 \frac{q^4}{4 m^2},
\end{equation}
where $\beta = 2 (U_3/(U_1 + U_3)) \cos^2{\theta}$ is a dimensionless small parameter.
The dispersion is linear
at the smallest $q$ and quadratic at larger $q$.
The crossover is at $q_c \sim (4\beta \mu_0 m c^2/(c-1))^{1/2} \sim \beta^{1/2} k_F$, which is parametrically smaller than $k_F$.
Correspondingly, the dynamical magnon susceptibility at $q < q_c$ is
\begin{equation}
    \chi_{\perp} (q, \Omega) =
    \frac{4 \mu^2 c^2}{\beta \mu_0 (c-1) q^2/m - \Omega^2},
    \label{eq:chi-perp}
\end{equation}
while
at
$q >q_c$, the poles are at $\Omega = \pm (q^2/(2m))  (c-1)/c$ and the spin susceptibility is the sum of two terms --- one proportional to $1/(\Omega  + (q^2/(2m))  (c-1)/c)$
and the other to $1/(\Omega  - (q^2/(2m))  (c-1)/c)$.
These terms describe two Goldstone modes of an SU(2) FM, moving in different directions.
The modes coming from  $D_- =0$ are out-of-phase fluctuations of magnetization in $K$ and $K'$ valleys.
These are gapped modes with the dispersion
$\Omega^2 =  32 \mu^2_0 c^2 ( U_3/(U_1+U_3))^2 \sin^2 {\theta} + O(q^2)$.

\section{Magnon-mediated 4-fermion interactions}
\label{sec:magnon-interaction}
\begin{figure}[htb]
    \centering
    \includegraphics[width=\linewidth]{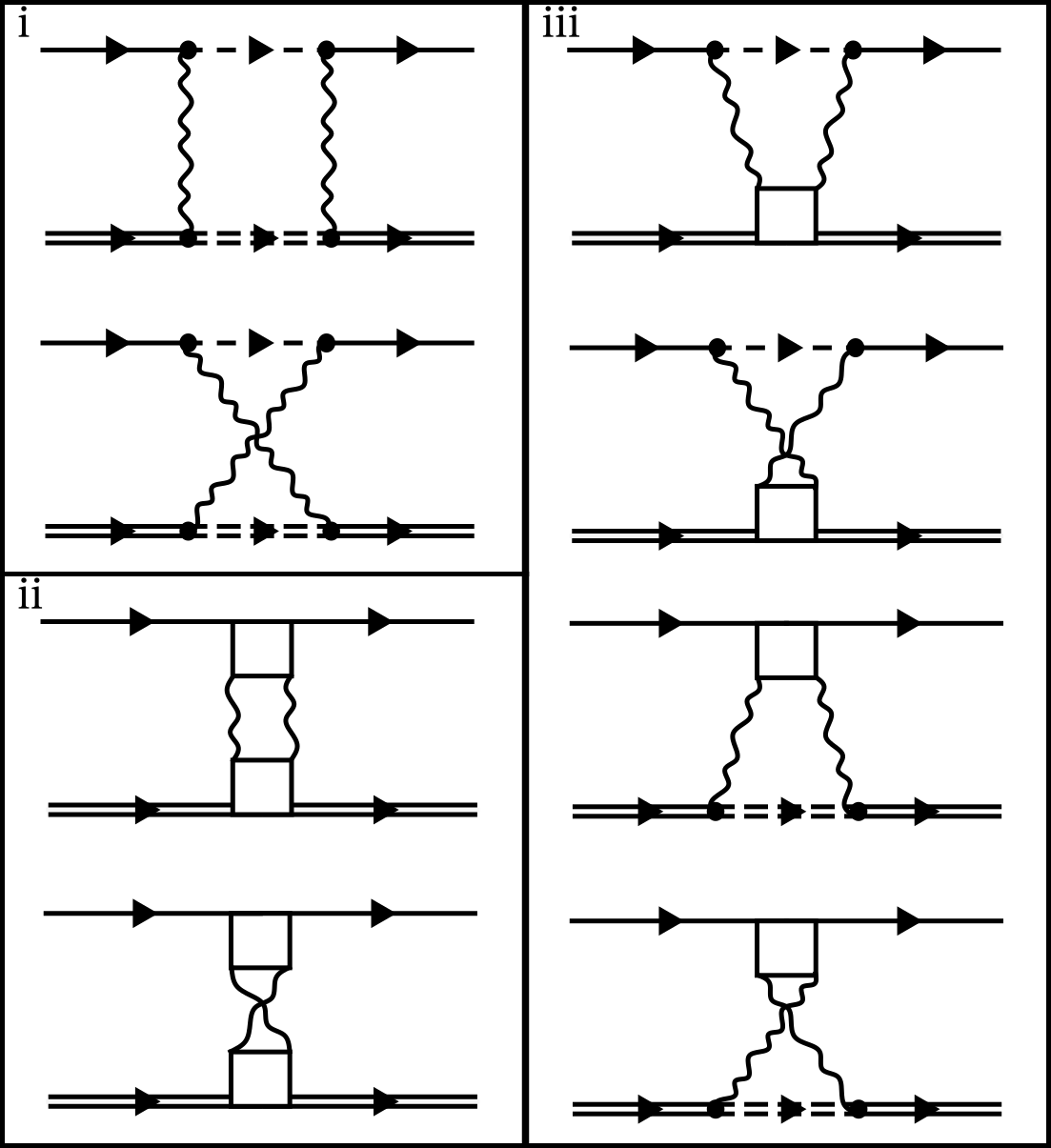}
    \caption{First- and second-order two-magnon
        processes contributing to
        the effective magnon-mediated pairing interaction between low-energy fermions.\label{fig:two-magnon-diagrams}}
\end{figure}

Our goal is to obtain the effective 4-fermion pairing interaction with zero total incoming momentum between low-energy fermions $f$ and $\tilde{f}$,  mediated by Goldstone magnons, check whether or not it is attractive, and estimate its strength.
A simple experimentation shows that
one can get a pairing interaction
between $f$ and ${\tilde f}$  fermions
by
processes involving two magnon propagators.
There are three such types of process (\cref{fig:two-magnon-diagrams}):  (i) second-order processes involving
two
4-fermion interactions mediated by a single magnon,
(ii) first-order processes involving a four-fermion interaction mediated by two magnons, and (iii) second-order processes involving one two-magnon vertex and two one-magnon vertices.
\begin{figure}[htb]
    \centering
    \includegraphics[width=\linewidth]{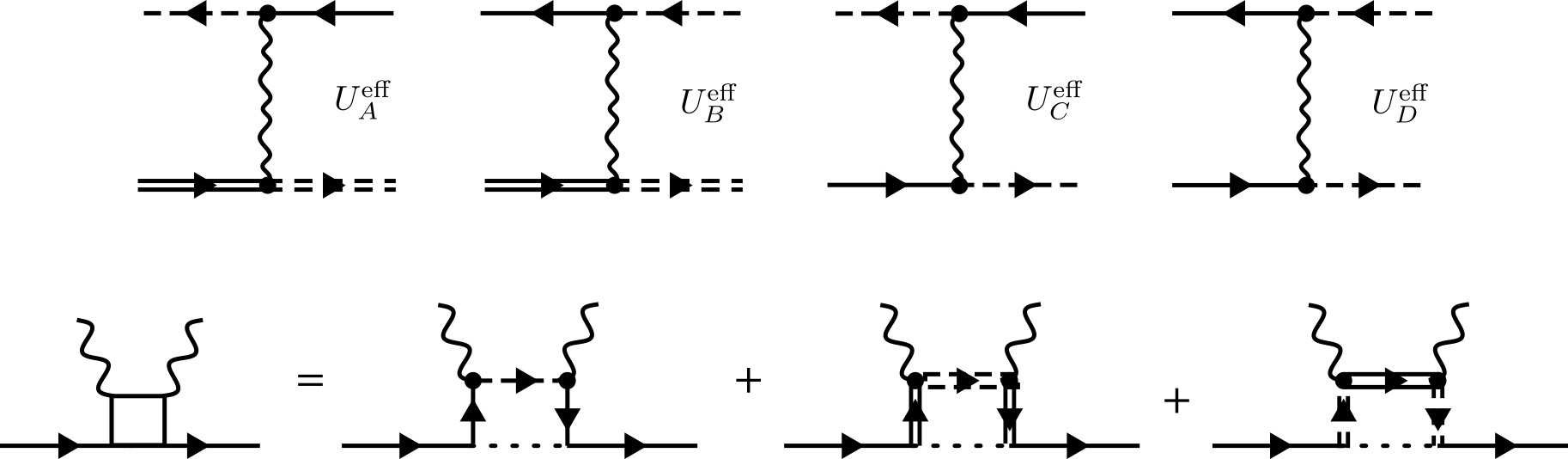}
    \caption{Top: Dynamical interactions $U^\text{eff}_{A-D}$ mediated by a single magnon.
        The interactions $U^\text{eff}_{A,C}$ are enabled by spin-orbit coupling, while the interactions $U^\text{eff}_{B,D}$ remain even in the SU(2) FM.
        Bottom: Diagrams comprising the scattering of two low-energy fermions and two magnons, involving all four of $U^\text{eff}_{A-D}$.
    }
    \label{fig:two-magnon-scattering}
\end{figure}

The building blocks are
the effective vertex between two low-energy fermions and two magnons, \cref{fig:two-magnon-scattering},
and
four effective interactions mediated by a single magnon  $U^\text{eff}_{A-D}$, \cref{fig:two-magnon-scattering} ($U^\text{eff}_{A-B}$  contribute to second-order one-magnon processes, $U^\text{eff}_{C-D}$ are parts of
first-order two-magnon process).
The effective two-fermion/two-magnon vertices
are the convolutions of three fermionic propagators, two of
spin-up fermions
and one of spin-down
fermion,
and
either the $U_1$ interaction or the rotated $U_3$ interaction from~\cref{eq:HU3}.
Each single-magnon vertex is
obtained by summing up ladder series,
where for each segment the rails are
one gapless propagator of $f$ or ${\tilde f}$ and one gapped propagator of $e$ or ${\tilde e}$ fermions and each rung is either the $U_1$ interaction or the rotated $U_3$ interaction from \cref{eq:HU3}.
We show the first few ladder diagrams for $U^\text{eff}_{C-D}$ in~\cref{sec:Ueff}.

Collecting the ladder series
in the same manner as was done in Ref.~\onlinecite{Dong2023b},
we find that each effective interaction is proportional to the susceptibility of a Goldstone magnon $\chi (q, \Omega)$.
The expressions for $U^\text{eff}_{A-D}$ for arbitrary $q$ are rather cumbersome and we present them in~\cref{sec:Ueff}.
At small $q <q_c$,
relevant for our calculations, these expressions simplify to
\begin{equation}
    U^\text{eff}_{B,D} = - U^\text{eff}_{A,C} =  U_3 \cos^2{\theta} \chi_{\perp} (q, \Omega).
\end{equation}
Similar expressions for $U^\text{eff}_{A,B}$ have been obtained in
Ref.~[\onlinecite{Dong2024}] for the model with spin-spin interaction between fermions.
Observe that the prefactor in~\cref{eq:chi-perp} does not vanish at $\Omega_m = q=0$.
As the authors of~\onlinecite{Dong2024} explained, this does not contradict the Adler principle for Goldstone bosons because $U^\text{eff}_{A-D}$ are ``non-diagonal'' elements which involve
both low-energy spin-up and finite energy spin-down fermions.
As we show immediately below, the full magnon-mediated interaction between low-energy
spin-up
fermions does satisfy the Adler principle.

\section{Magnon-mediated pairing interaction}
\label{sec:pairing-interaction}
We are now in a position to obtain the full magnon-mediated pairing interaction  with zero total momentum between low-energy spin-up fermions, $\Gamma^{sc}_{2mag}  (k,-k;k +\delta,-k-\delta) = \Gamma^{sc}_{2mag} (\delta)$.
As we said, we assume spatially-even, valley isospin-odd gap symmetry for equal spin pairing.
For such pairing symmetry, the sign of the pairing interaction can be determined with high confidence by evaluating $\Gamma^{sc}_{2mag} (0)$.
This is what we will do.

\begin{figure}
    \includegraphics[width=\linewidth]{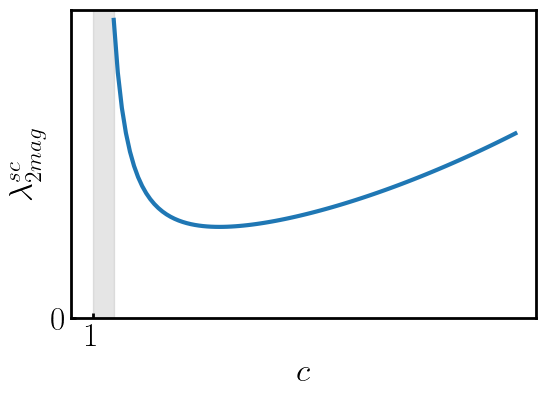}
    \caption{Qualitative behavior of the dimensionless pairing coupling $\lambda^{sc}_{2mag}$ as a function of the dimensionless interaction strength $c$.
        \label{fig:lambda_qual}
    }
\end{figure}

The effective magnon-mediated pairing interaction $\Gamma^{sc}_{2mag} (0)$ is the sum of the contributions from all three types of diagrams in~\cref{fig:two-magnon-diagrams}.
Collecting these contributions, we obtain on the Matsubara axis,
\begin{equation}
    \Gamma^{sc}_{2mag} (0) =
    \int \frac{d^2 q d \Omega_m}{(2\pi)^3}  A (q, \Omega_m) \chi^2_{\perp} (q, \Omega_m)
    \label{eq:Usc}
\end{equation} where  $\chi_{\perp} (q, \Omega_m)$ is given by~\cref{eq:chi-perp} with
$\Omega \to i \Omega_m$ and
$A(q, \Omega_m)$  is the combined contribution from first and second order two-magnon scattering processes.
We find $A(0,0) =0$, i.e., the vertex function for the interaction between low-energy fermions, mediated by Goldstone magnons, vanishes
at the ordering momentum and zero frequency~\footnote{We are thankful to Erez Berg for emphasizing that $A(0,0)$ must vanish.}
This is the manifestation of the Adler principle in a FM.
Expanding further $A(q, \Omega_m)$ at the smallest $q < q_c = \beta^{1/2} k_F$ and $|\Omega_m| < \Omega_c = q^2_c/(2m) =\beta \mu_0$, we find that the $q^2$ and $\Omega^2_m$ terms also vanish.
At fourth order, we obtain
\begin{equation}
    A(q, \Omega_m) = -  \frac{U^2_3 \cos^4{\theta}}{(2\mu^2_0 c^2)^3} {\bar A} (q, \Omega_m),
\end{equation}
where
\begin{equation}
    \bar{A} (q, \Omega_m) = \Omega^4_m + \frac{q^2 \mu_0}{m} \Omega^2_m (c+1) + \left(\frac{q^2 \mu_0}{2m}\right)^2 (c-1)^2.
    \label{eq:Akernel}
\end{equation}
We see that
$A(q, \Omega_m) <0$, i.e., $\Gamma^{sc}_{2mag} (0)$ is negative, i.e., \emph{attractive for the pairing}.
We emphasize that the sign of $\Gamma^{sc}_{2mag} (0)$ could not be established without the actual computation of $A (q,\Omega_m)$.
Extending calculations to a finite momentum and frequency transfer between incoming and outgoing fermions, we find that $\Gamma^{sc}_{2mag} (\delta)$ remains negative at a non-zero $\delta$ and drops at $\delta \sim q_c$ and at frequency transfer of order $\Omega_c$. Substituting~\cref{eq:Akernel} into~\cref{eq:Usc} and restricting the integration over $q$ to $q \leq  q_c$ and $\Omega_m < \Omega_c$, where~\cref{eq:Akernel,eq:chi-perp} hold, we find
\begin{equation}
    N_F \Gamma^\text{sc}_{2mag} (0) \sim  \frac{c^3}{c-1} \beta^2.
\end{equation}
We emphasize that this contribution is entirely due to SOC.
Substituting
this interaction into the gap equation we obtain after standard manipulations the dimensionless coupling constant
\begin{equation}
    \lambda^\text{sc}_{2mag} a\frac{c^4}{(c-1)^{3/2}} \beta^{5/2},
    \label{eq:lambdaSOC-estimate}
\end{equation}
where $a=O(1)$ is a numerical factor.
The coupling is weak in $\beta$, but it gets enhanced near the onset of a FM state, where $c \approx 1$, and  deep inside a CFM, where $c \gg 1$.
Note that our theory is only valid for $q_c \ll k_F$, so there is a natural cutoff $c-1 > \beta$.
We thus expect the coupling to saturate at a scale $O(\beta)$ as the critical point is approached.
We plot $\lambda^\text{sc}_{2mag}$ in \cref{fig:lambda_qual}.

There also exists a contribution to $\Gamma^{sc}_{2mag} (\delta)$  from larger $q > q_c$ and $\Omega_m > \Omega_c$,
where the magnon dispersion is quadratic. We find that this last contribution is essentially static, scales as $U^2_3 N_F$
This term is a
sub-leading
correction to
a bare repulsive interaction $U_2$.
Together, the $U_2$ term and the contribution from $q>q_c$ form a static repulsive interaction, which holds already at energies of order $\mu_0$.
In contrast, the SOC-induced pairing interaction exists at energies smaller than $\Omega_c \sim\beta \mu_0$, where, we remind, $\beta$ is a small parameter in our consideration.
In this respect, the situation is similar to that for a system with an attractive electron-phonon interaction confined to low energies and a stronger repulsive Coulomb (or Hubbard) interaction (see. e.g. Ref. [\onlinecite{Coleman2015}]).
We conjecture that, like there,  superconductivity develops due to retardation, and  $T_c \sim \beta \mu_0 e^{-1/(|\lambda^{sc}_{2mag}|-\mu^*)}$, where $\mu^*$ is the renormalized contribution from the static repulsion.
More sophisticated calculations are needed to obtain $\mu^*$
and verify whether this mechanism can explain superconductivity
observed in the CFM state of BBG and RTG placed next to
WSe$_2$~\cite{Holleis2023,Zhang2023a,Patterson2024}.
Still, the observation that SC holds only for carriers subject to SOC
is fully consistent with our theoretical finding.
We also emphasize that our $\lambda^{sc}_{2mag}$, while generally small, gets substantially enhanced near the
onset of a CFM order and deep inside the CFM phase.
The enhancement of the attraction near a CFM boundary is consistent with recent
observation of superconductivity with $T_c \sim \SI{300}{mK}$  near the onset of CFM in RTG~\cite{Patterson2024}.
The enhancement of the attraction deep inside a FM  phase is consistent with the experiments on BBG next to WSe$_2$~\cite{Holleis2023},
which detected superconductivity in a parameter range well inside a half-metal state, which we interpret as a CFM.

Before concluding, we briefly compare our results with the ones in Ref.~\onlinecite{Dong2024}.
First, our expressions for the magnon propagator and  one-magnon interactions $U^\text{eff}_{A,B}$
fully agree with [\onlinecite{Dong2024}], when re-expressed in terms of the corresponding parameters (the ratio $U_3/U_1$ in our model plays the same role as the ratio of two spin-spin interactions, $J/V$, in their model).
Second, the authors of [\onlinecite{Dong2024}] considered the case when there exists a Fermi surface of minority carriers.
They analyzed the pairing at a momentum transfer $\delta k$,  at which both spin-up and spin-down fermions are near their respective Fermi surfaces, and solved the $2\times 2$ gap equation similar to how this has been done for Fe-based superconductors, by using $U^\text{eff}_A$ as an inter-pocket pair-hopping interaction.
They found a larger $\lambda^\text{sc}_{A} = O(1)$  with no dependence on $\beta$.
This result, however, holds if $\delta k$ is smaller than $q_c \sim \beta^{1/2} k_F$, i.e., when the Fermi surfaces of majority of minority carriers are of about the same size.
Within our model, there is no Fermi surface for minority carriers and the attraction comes from virtual processes involving gapped states of minority carriers.
Whether or not a minority Fermi surface exists in the parameter range where superconductivity with the highest $T_c$ has been discovered, i.e., whether the normal state above $T_c$ is a true half-metal or ``almost'' half-metal, will almost certainly be settled by the experiments.
We note in passing that within our model, a CFM state with a small Fermi surface for minority carriers does appear in some portion of the FM-ordered phase when trigonal warping is included.
However, the same trigonal warping then has to be included into the analysis of fermion-magnon interaction
\footnote{The authors of [\onlinecite{Dong2024}] also speculated that although the theoretical $\beta$ contains a small ratio ($U_3/U_1$ in our case), it may be $O(1)$ for the actual parameters.
    In this situation, their $\delta k \sim k_F$, as is generally expected, and also our $\lambda^{sc}_{2mag}$ is not reduced by $\beta$.
    However, for $\beta = O(1)$, there is no clear distinction between ``low-energy'' and ``high-energy'' contributions and a repulsive $U_2$ becomes more detrimental for superconductivity.}.

\section{Conclusion}
\label{sec:conclusion}
In this communication, we analyzed the origin of superconductivity in a half-metal state in BBG and RTG placed in proximity to WSe$_2$, by modeling these materials as a two-valley system of interacting fermions in the presence of an Ising spin-orbit coupling with opposite sign in the two valleys.
We obtained a CFM order, which gives rise to a half-metal with no fermions in two out of four bands,  obtained the spectrum of Goldstone excitations and derived spin-flip 4-fermion interactions mediated by a single magnon and spin-preserving interactions mediated by two magnons.
We argued that both types of processes have to be included
into the effective pairing interaction between low-energy fermions from the filled bands, mediated by two Goldstone magnons.
We derived this interaction for valley-odd/spatially-even order parameter and verified that the fermion-magnon vertex function vanishes in the limit of zero magnon momentum and frequency, in agreement with the Adler principle for Goldstone bosons.
Expanding in momentum and frequency, we obtained an attractive dynamical interaction, which is induced by SOC and confined to small energies, where the magnon dispersion is linear.
We argued that due to retardation this interaction may give rise to superconductivity even though it is smaller than a static repulsion, in close analogy with how phonon-mediated attraction gives rise to pairing in the presence of a Coulomb (Hubbard) repulsion.

Two final remarks: (i) In this work we considered a $q=0$ ferromagnetic order. The analysis done here can be straightforwardly extended to the case of a magnetic order with momentum $K-K'$ (a spin inter-valley order).
(ii)  We discussed  pairing with zero total momentum, which involves fermions from both valleys.
Another possibility is a pairing between fermions within a valley --- a PDW with momentum $2K$ or $2K'$
(Refs.~[\onlinecite{
    Chou2025,*Geier2024,*Yang2024,*Qin2024,*Jahin2025,*Gil2025,*Dong2025,*Gaggioli2025,*Yang2024, *Parramartinez2025,*Kim2025,*Christos2025}]).
Such a pairing exists even if $U_3 =\lambda =0$, and the corresponding coupling constant is $O(1)$ parameter-wise (Ref.~[\onlinecite{Raines2025a}]).
For a circular Fermi surface, such a pairing (with a spatially odd gap, as only one valley is involved) is a strong competitor to zero momentum pairing.
However, for a non-circular Fermi surface, a finite momentum pairing is less favorable, at least at a moderate coupling, because there is no Cooper logarithm for a PDW.

\begin{acknowledgments}
We acknowledge with thanks useful discussions with Jason Alicea, Erez Berg, Hart Goldman, Alex Kamenev, Patrick Lee, Stevan Nadj-Perge, Pavel Nosov, Matthias Scheurer, Alex Thompson, Andrea Young and particularly Zhiyu Dong.
The work was supported by the U.S.\@ Department of Energy, Office of Science, Basic Energy Sciences, under Award No.~DE-SC0014402.
\end{acknowledgments}

\bibliography{references_2v}

\begin{thebibliography}{55}%
\makeatletter
\providecommand \@ifxundefined [1]{%
 \@ifx{#1\undefined}
}%
\providecommand \@ifnum [1]{%
 \ifnum #1\expandafter \@firstoftwo
 \else \expandafter \@secondoftwo
 \fi
}%
\providecommand \@ifx [1]{%
 \ifx #1\expandafter \@firstoftwo
 \else \expandafter \@secondoftwo
 \fi
}%
\providecommand \natexlab [1]{#1}%
\providecommand \enquote  [1]{``#1''}%
\providecommand \bibnamefont  [1]{#1}%
\providecommand \bibfnamefont [1]{#1}%
\providecommand \citenamefont [1]{#1}%
\providecommand \href@noop [0]{\@secondoftwo}%
\providecommand \href [0]{\begingroup \@sanitize@url \@href}%
\providecommand \@href[1]{\@@startlink{#1}\@@href}%
\providecommand \@@href[1]{\endgroup#1\@@endlink}%
\providecommand \@sanitize@url [0]{\catcode `\\12\catcode `\$12\catcode
  `\&12\catcode `\#12\catcode `\^12\catcode `\_12\catcode `\%12\relax}%
\providecommand \@@startlink[1]{}%
\providecommand \@@endlink[0]{}%
\providecommand \url  [0]{\begingroup\@sanitize@url \@url }%
\providecommand \@url [1]{\endgroup\@href {#1}{\urlprefix }}%
\providecommand \urlprefix  [0]{URL }%
\providecommand \Eprint [0]{\href }%
\providecommand \doibase [0]{https://doi.org/}%
\providecommand \selectlanguage [0]{\@gobble}%
\providecommand \bibinfo  [0]{\@secondoftwo}%
\providecommand \bibfield  [0]{\@secondoftwo}%
\providecommand \translation [1]{[#1]}%
\providecommand \BibitemOpen [0]{}%
\providecommand \bibitemStop [0]{}%
\providecommand \bibitemNoStop [0]{.\EOS\space}%
\providecommand \EOS [0]{\spacefactor3000\relax}%
\providecommand \BibitemShut  [1]{\csname bibitem#1\endcsname}%
\let\auto@bib@innerbib\@empty
\bibitem [{\citenamefont {Cao}\ \emph {et~al.}(2018)\citenamefont {Cao},
  \citenamefont {Fatemi}, \citenamefont {Fang}, \citenamefont {Watanabe},
  \citenamefont {Taniguchi}, \citenamefont {Kaxiras},\ and\ \citenamefont
  {{Jarillo-Herrero}}}]{Cao2018a}%
  \BibitemOpen
  \bibfield  {author} {\bibinfo {author} {\bibfnamefont {Y.}~\bibnamefont
  {Cao}}, \bibinfo {author} {\bibfnamefont {V.}~\bibnamefont {Fatemi}},
  \bibinfo {author} {\bibfnamefont {S.}~\bibnamefont {Fang}}, \bibinfo {author}
  {\bibfnamefont {K.}~\bibnamefont {Watanabe}}, \bibinfo {author}
  {\bibfnamefont {T.}~\bibnamefont {Taniguchi}}, \bibinfo {author}
  {\bibfnamefont {E.}~\bibnamefont {Kaxiras}},\ and\ \bibinfo {author}
  {\bibfnamefont {P.}~\bibnamefont {{Jarillo-Herrero}}},\ }\bibfield  {title}
  {\bibinfo {title} {Unconventional superconductivity in magic-angle graphene
  superlattices},\ }\href {https://doi.org/10.1038/nature26160} {\bibfield
  {journal} {\bibinfo  {journal} {Nature}\ }\textbf {\bibinfo {volume} {556}},\
  \bibinfo {pages} {43} (\bibinfo {year} {2018})}\BibitemShut {NoStop}%
\bibitem [{\citenamefont {Oh}\ \emph {et~al.}(2021)\citenamefont {Oh},
  \citenamefont {Nuckolls}, \citenamefont {Wong}, \citenamefont {Lee},
  \citenamefont {Liu}, \citenamefont {Watanabe}, \citenamefont {Taniguchi},\
  and\ \citenamefont {Yazdani}}]{Oh2021}%
  \BibitemOpen
  \bibfield  {author} {\bibinfo {author} {\bibfnamefont {M.}~\bibnamefont
  {Oh}}, \bibinfo {author} {\bibfnamefont {K.~P.}\ \bibnamefont {Nuckolls}},
  \bibinfo {author} {\bibfnamefont {D.}~\bibnamefont {Wong}}, \bibinfo {author}
  {\bibfnamefont {R.~L.}\ \bibnamefont {Lee}}, \bibinfo {author} {\bibfnamefont
  {X.}~\bibnamefont {Liu}}, \bibinfo {author} {\bibfnamefont {K.}~\bibnamefont
  {Watanabe}}, \bibinfo {author} {\bibfnamefont {T.}~\bibnamefont
  {Taniguchi}},\ and\ \bibinfo {author} {\bibfnamefont {A.}~\bibnamefont
  {Yazdani}},\ }\bibfield  {title} {\bibinfo {title} {Evidence for
  unconventional superconductivity in twisted bilayer graphene},\ }\href
  {https://doi.org/10.1038/s41586-021-04121-x} {\bibfield  {journal} {\bibinfo
  {journal} {Nature}\ }\textbf {\bibinfo {volume} {600}},\ \bibinfo {pages}
  {240} (\bibinfo {year} {2021})}\BibitemShut {NoStop}%
\bibitem [{\citenamefont {Zhou}\ \emph {et~al.}(2022)\citenamefont {Zhou},
  \citenamefont {Holleis}, \citenamefont {Saito}, \citenamefont {Cohen},
  \citenamefont {Huynh}, \citenamefont {Patterson}, \citenamefont {Yang},
  \citenamefont {Taniguchi}, \citenamefont {Watanabe},\ and\ \citenamefont
  {Young}}]{Zhou2022a}%
  \BibitemOpen
  \bibfield  {author} {\bibinfo {author} {\bibfnamefont {H.}~\bibnamefont
  {Zhou}}, \bibinfo {author} {\bibfnamefont {L.}~\bibnamefont {Holleis}},
  \bibinfo {author} {\bibfnamefont {Y.}~\bibnamefont {Saito}}, \bibinfo
  {author} {\bibfnamefont {L.}~\bibnamefont {Cohen}}, \bibinfo {author}
  {\bibfnamefont {W.}~\bibnamefont {Huynh}}, \bibinfo {author} {\bibfnamefont
  {C.~L.}\ \bibnamefont {Patterson}}, \bibinfo {author} {\bibfnamefont
  {F.}~\bibnamefont {Yang}}, \bibinfo {author} {\bibfnamefont {T.}~\bibnamefont
  {Taniguchi}}, \bibinfo {author} {\bibfnamefont {K.}~\bibnamefont
  {Watanabe}},\ and\ \bibinfo {author} {\bibfnamefont {A.~F.}\ \bibnamefont
  {Young}},\ }\bibfield  {title} {\bibinfo {title} {Isospin magnetism and
  spin-polarized superconductivity in {{Bernal}} bilayer graphene},\ }\href
  {https://doi.org/10.1126/science.abm8386} {\bibfield  {journal} {\bibinfo
  {journal} {Science}\ }\textbf {\bibinfo {volume} {375}},\ \bibinfo {pages}
  {774} (\bibinfo {year} {2022})}\BibitemShut {NoStop}%
\bibitem [{\citenamefont {Zhang}\ \emph {et~al.}(2023)\citenamefont {Zhang},
  \citenamefont {Polski}, \citenamefont {Thomson}, \citenamefont
  {{Lantagne-Hurtubise}}, \citenamefont {Lewandowski}, \citenamefont {Zhou},
  \citenamefont {Watanabe}, \citenamefont {Taniguchi}, \citenamefont {Alicea},\
  and\ \citenamefont {{Nadj-Perge}}}]{Zhang2023a}%
  \BibitemOpen
  \bibfield  {author} {\bibinfo {author} {\bibfnamefont {Y.}~\bibnamefont
  {Zhang}}, \bibinfo {author} {\bibfnamefont {R.}~\bibnamefont {Polski}},
  \bibinfo {author} {\bibfnamefont {A.}~\bibnamefont {Thomson}}, \bibinfo
  {author} {\bibfnamefont {{\'E}.}~\bibnamefont {{Lantagne-Hurtubise}}},
  \bibinfo {author} {\bibfnamefont {C.}~\bibnamefont {Lewandowski}}, \bibinfo
  {author} {\bibfnamefont {H.}~\bibnamefont {Zhou}}, \bibinfo {author}
  {\bibfnamefont {K.}~\bibnamefont {Watanabe}}, \bibinfo {author}
  {\bibfnamefont {T.}~\bibnamefont {Taniguchi}}, \bibinfo {author}
  {\bibfnamefont {J.}~\bibnamefont {Alicea}},\ and\ \bibinfo {author}
  {\bibfnamefont {S.}~\bibnamefont {{Nadj-Perge}}},\ }\bibfield  {title}
  {\bibinfo {title} {Enhanced superconductivity in spin--orbit proximitized
  bilayer graphene},\ }\href {https://doi.org/10.1038/s41586-022-05446-x}
  {\bibfield  {journal} {\bibinfo  {journal} {Nature}\ }\textbf {\bibinfo
  {volume} {613}},\ \bibinfo {pages} {268} (\bibinfo {year}
  {2023})}\BibitemShut {NoStop}%
\bibitem [{\citenamefont {Holleis}\ \emph {et~al.}(2023)\citenamefont
  {Holleis}, \citenamefont {Patterson}, \citenamefont {Zhang}, \citenamefont
  {Yoo}, \citenamefont {Zhou}, \citenamefont {Taniguchi}, \citenamefont
  {Watanabe}, \citenamefont {{Nadj-Perge}},\ and\ \citenamefont
  {Young}}]{Holleis2023}%
  \BibitemOpen
  \bibfield  {author} {\bibinfo {author} {\bibfnamefont {L.}~\bibnamefont
  {Holleis}}, \bibinfo {author} {\bibfnamefont {C.~L.}\ \bibnamefont
  {Patterson}}, \bibinfo {author} {\bibfnamefont {Y.}~\bibnamefont {Zhang}},
  \bibinfo {author} {\bibfnamefont {H.~M.}\ \bibnamefont {Yoo}}, \bibinfo
  {author} {\bibfnamefont {H.}~\bibnamefont {Zhou}}, \bibinfo {author}
  {\bibfnamefont {T.}~\bibnamefont {Taniguchi}}, \bibinfo {author}
  {\bibfnamefont {K.}~\bibnamefont {Watanabe}}, \bibinfo {author}
  {\bibfnamefont {S.}~\bibnamefont {{Nadj-Perge}}},\ and\ \bibinfo {author}
  {\bibfnamefont {A.~F.}\ \bibnamefont {Young}},\ }\href
  {https://doi.org/10.48550/arXiv.2303.00742} {\bibinfo {title} {Ising
  {{Superconductivity}} and {{Nematicity}} in {{Bernal Bilayer Graphene}} with
  {{Strong Spin Orbit Coupling}}}} (\bibinfo {year} {2023}),\ \Eprint
  {https://arxiv.org/abs/2303.00742} {arXiv:2303.00742 [cond-mat.supr-con]}
  \BibitemShut {NoStop}%
\bibitem [{\citenamefont {Zhou}\ \emph {et~al.}(2021)\citenamefont {Zhou},
  \citenamefont {Xie}, \citenamefont {Taniguchi}, \citenamefont {Watanabe},\
  and\ \citenamefont {Young}}]{Zhou2021a}%
  \BibitemOpen
  \bibfield  {author} {\bibinfo {author} {\bibfnamefont {H.}~\bibnamefont
  {Zhou}}, \bibinfo {author} {\bibfnamefont {T.}~\bibnamefont {Xie}}, \bibinfo
  {author} {\bibfnamefont {T.}~\bibnamefont {Taniguchi}}, \bibinfo {author}
  {\bibfnamefont {K.}~\bibnamefont {Watanabe}},\ and\ \bibinfo {author}
  {\bibfnamefont {A.~F.}\ \bibnamefont {Young}},\ }\bibfield  {title} {\bibinfo
  {title} {Superconductivity in rhombohedral trilayer graphene},\ }\href
  {https://doi.org/10.1038/s41586-021-03926-0} {\bibfield  {journal} {\bibinfo
  {journal} {Nature}\ }\textbf {\bibinfo {volume} {598}},\ \bibinfo {pages}
  {434} (\bibinfo {year} {2021})}\BibitemShut {NoStop}%
\bibitem [{\citenamefont {Han}\ \emph {et~al.}(2025)\citenamefont {Han},
  \citenamefont {Lu}, \citenamefont {Hadjri}, \citenamefont {Shi},
  \citenamefont {Wu}, \citenamefont {Xu}, \citenamefont {Yao}, \citenamefont
  {Cotten}, \citenamefont {Sedeh}, \citenamefont {Weldeyesus}, \citenamefont
  {Yang}, \citenamefont {Seo}, \citenamefont {Ye}, \citenamefont {Zhou},
  \citenamefont {Liu}, \citenamefont {Shi}, \citenamefont {Hua}, \citenamefont
  {Watanabe}, \citenamefont {Taniguchi}, \citenamefont {Xiong}, \citenamefont
  {Zumb{\"u}hl}, \citenamefont {Fu},\ and\ \citenamefont {Ju}}]{Han2025}%
  \BibitemOpen
  \bibfield  {author} {\bibinfo {author} {\bibfnamefont {T.}~\bibnamefont
  {Han}}, \bibinfo {author} {\bibfnamefont {Z.}~\bibnamefont {Lu}}, \bibinfo
  {author} {\bibfnamefont {Z.}~\bibnamefont {Hadjri}}, \bibinfo {author}
  {\bibfnamefont {L.}~\bibnamefont {Shi}}, \bibinfo {author} {\bibfnamefont
  {Z.}~\bibnamefont {Wu}}, \bibinfo {author} {\bibfnamefont {W.}~\bibnamefont
  {Xu}}, \bibinfo {author} {\bibfnamefont {Y.}~\bibnamefont {Yao}}, \bibinfo
  {author} {\bibfnamefont {A.~A.}\ \bibnamefont {Cotten}}, \bibinfo {author}
  {\bibfnamefont {O.~S.}\ \bibnamefont {Sedeh}}, \bibinfo {author}
  {\bibfnamefont {H.}~\bibnamefont {Weldeyesus}}, \bibinfo {author}
  {\bibfnamefont {J.}~\bibnamefont {Yang}}, \bibinfo {author} {\bibfnamefont
  {J.}~\bibnamefont {Seo}}, \bibinfo {author} {\bibfnamefont {S.}~\bibnamefont
  {Ye}}, \bibinfo {author} {\bibfnamefont {M.}~\bibnamefont {Zhou}}, \bibinfo
  {author} {\bibfnamefont {H.}~\bibnamefont {Liu}}, \bibinfo {author}
  {\bibfnamefont {G.}~\bibnamefont {Shi}}, \bibinfo {author} {\bibfnamefont
  {Z.}~\bibnamefont {Hua}}, \bibinfo {author} {\bibfnamefont {K.}~\bibnamefont
  {Watanabe}}, \bibinfo {author} {\bibfnamefont {T.}~\bibnamefont {Taniguchi}},
  \bibinfo {author} {\bibfnamefont {P.}~\bibnamefont {Xiong}}, \bibinfo
  {author} {\bibfnamefont {D.~M.}\ \bibnamefont {Zumb{\"u}hl}}, \bibinfo
  {author} {\bibfnamefont {L.}~\bibnamefont {Fu}},\ and\ \bibinfo {author}
  {\bibfnamefont {L.}~\bibnamefont {Ju}},\ }\bibfield  {title} {\bibinfo
  {title} {Signatures of chiral superconductivity in rhombohedral graphene},\
  }\href {https://doi.org/10.1038/s41586-025-09169-7} {\bibfield  {journal}
  {\bibinfo  {journal} {Nature}\ ,\ \bibinfo {pages} {1}} (\bibinfo {year}
  {2025})}\BibitemShut {NoStop}%
\bibitem [{\citenamefont {Wang}\ \emph {et~al.}(2016)\citenamefont {Wang},
  \citenamefont {Ki}, \citenamefont {Khoo}, \citenamefont {Mauro},
  \citenamefont {Berger}, \citenamefont {Levitov},\ and\ \citenamefont
  {Morpurgo}}]{Wang2016}%
  \BibitemOpen
  \bibfield  {author} {\bibinfo {author} {\bibfnamefont {Z.}~\bibnamefont
  {Wang}}, \bibinfo {author} {\bibfnamefont {D.-K.}\ \bibnamefont {Ki}},
  \bibinfo {author} {\bibfnamefont {J.~Y.}\ \bibnamefont {Khoo}}, \bibinfo
  {author} {\bibfnamefont {D.}~\bibnamefont {Mauro}}, \bibinfo {author}
  {\bibfnamefont {H.}~\bibnamefont {Berger}}, \bibinfo {author} {\bibfnamefont
  {L.~S.}\ \bibnamefont {Levitov}},\ and\ \bibinfo {author} {\bibfnamefont
  {A.~F.}\ \bibnamefont {Morpurgo}},\ }\bibfield  {title} {\bibinfo {title}
  {Origin and {{Magnitude}} of `{{Designer}}' {{Spin-Orbit Interaction}} in
  {{Graphene}} on {{Semiconducting Transition Metal Dichalcogenides}}},\ }\href
  {https://doi.org/10.1103/physrevx.6.041020} {\bibfield  {journal} {\bibinfo
  {journal} {Phys. Rev. X}\ }\textbf {\bibinfo {volume} {6}},\ \bibinfo {pages}
  {041020} (\bibinfo {year} {2016})}\BibitemShut {NoStop}%
\bibitem [{\citenamefont {Island}\ \emph {et~al.}(2019)\citenamefont {Island},
  \citenamefont {Cui}, \citenamefont {Lewandowski}, \citenamefont {Khoo},
  \citenamefont {Spanton}, \citenamefont {Zhou}, \citenamefont {Rhodes},
  \citenamefont {Hone}, \citenamefont {Taniguchi}, \citenamefont {Watanabe},
  \citenamefont {Levitov}, \citenamefont {Zaletel},\ and\ \citenamefont
  {Young}}]{Island2019}%
  \BibitemOpen
  \bibfield  {author} {\bibinfo {author} {\bibfnamefont {J.~O.}\ \bibnamefont
  {Island}}, \bibinfo {author} {\bibfnamefont {X.}~\bibnamefont {Cui}},
  \bibinfo {author} {\bibfnamefont {C.}~\bibnamefont {Lewandowski}}, \bibinfo
  {author} {\bibfnamefont {J.~Y.}\ \bibnamefont {Khoo}}, \bibinfo {author}
  {\bibfnamefont {E.~M.}\ \bibnamefont {Spanton}}, \bibinfo {author}
  {\bibfnamefont {H.}~\bibnamefont {Zhou}}, \bibinfo {author} {\bibfnamefont
  {D.}~\bibnamefont {Rhodes}}, \bibinfo {author} {\bibfnamefont {J.~C.}\
  \bibnamefont {Hone}}, \bibinfo {author} {\bibfnamefont {T.}~\bibnamefont
  {Taniguchi}}, \bibinfo {author} {\bibfnamefont {K.}~\bibnamefont {Watanabe}},
  \bibinfo {author} {\bibfnamefont {L.~S.}\ \bibnamefont {Levitov}}, \bibinfo
  {author} {\bibfnamefont {M.~P.}\ \bibnamefont {Zaletel}},\ and\ \bibinfo
  {author} {\bibfnamefont {A.~F.}\ \bibnamefont {Young}},\ }\bibfield  {title}
  {\bibinfo {title} {Spin--orbit-driven band inversion in bilayer graphene by
  the van der {{Waals}} proximity effect},\ }\href
  {https://doi.org/10.1038/s41586-019-1304-2} {\bibfield  {journal} {\bibinfo
  {journal} {Nature}\ }\textbf {\bibinfo {volume} {571}},\ \bibinfo {pages}
  {85} (\bibinfo {year} {2019})}\BibitemShut {NoStop}%
\bibitem [{\citenamefont {Patterson}\ \emph {et~al.}(2024)\citenamefont
  {Patterson}, \citenamefont {Sheekey}, \citenamefont {Arp}, \citenamefont
  {Holleis}, \citenamefont {Koh}, \citenamefont {Choi}, \citenamefont {Xie},
  \citenamefont {Xu}, \citenamefont {Redekop}, \citenamefont {Babikyan},
  \citenamefont {Zhou}, \citenamefont {Cheng}, \citenamefont {Taniguchi},
  \citenamefont {Watanabe}, \citenamefont {Jin}, \citenamefont
  {{Lantagne-Hurtubise}}, \citenamefont {Alicea},\ and\ \citenamefont
  {Young}}]{Patterson2024}%
  \BibitemOpen
  \bibfield  {author} {\bibinfo {author} {\bibfnamefont {C.~L.}\ \bibnamefont
  {Patterson}}, \bibinfo {author} {\bibfnamefont {O.~I.}\ \bibnamefont
  {Sheekey}}, \bibinfo {author} {\bibfnamefont {T.~B.}\ \bibnamefont {Arp}},
  \bibinfo {author} {\bibfnamefont {L.~F.~W.}\ \bibnamefont {Holleis}},
  \bibinfo {author} {\bibfnamefont {J.~M.}\ \bibnamefont {Koh}}, \bibinfo
  {author} {\bibfnamefont {Y.}~\bibnamefont {Choi}}, \bibinfo {author}
  {\bibfnamefont {T.}~\bibnamefont {Xie}}, \bibinfo {author} {\bibfnamefont
  {S.}~\bibnamefont {Xu}}, \bibinfo {author} {\bibfnamefont {E.}~\bibnamefont
  {Redekop}}, \bibinfo {author} {\bibfnamefont {G.}~\bibnamefont {Babikyan}},
  \bibinfo {author} {\bibfnamefont {H.}~\bibnamefont {Zhou}}, \bibinfo {author}
  {\bibfnamefont {X.}~\bibnamefont {Cheng}}, \bibinfo {author} {\bibfnamefont
  {T.}~\bibnamefont {Taniguchi}}, \bibinfo {author} {\bibfnamefont
  {K.}~\bibnamefont {Watanabe}}, \bibinfo {author} {\bibfnamefont
  {C.}~\bibnamefont {Jin}}, \bibinfo {author} {\bibfnamefont {E.}~\bibnamefont
  {{Lantagne-Hurtubise}}}, \bibinfo {author} {\bibfnamefont {J.}~\bibnamefont
  {Alicea}},\ and\ \bibinfo {author} {\bibfnamefont {A.~F.}\ \bibnamefont
  {Young}},\ }\href@noop {} {\bibinfo {title} {Superconductivity and spin
  canting in spin-orbit proximitized rhombohedral trilayer graphene}} (\bibinfo
  {year} {2024}),\ \Eprint {https://arxiv.org/abs/2408.10190} {arXiv:2408.10190
  [cond-mat.mes-hall]} \BibitemShut {NoStop}%
\bibitem [{\citenamefont {Schrieffer}\ \emph {et~al.}(1989)\citenamefont
  {Schrieffer}, \citenamefont {Wen},\ and\ \citenamefont
  {Zhang}}]{Schrieffer1989}%
  \BibitemOpen
  \bibfield  {author} {\bibinfo {author} {\bibfnamefont {J.~R.}\ \bibnamefont
  {Schrieffer}}, \bibinfo {author} {\bibfnamefont {X.~G.}\ \bibnamefont
  {Wen}},\ and\ \bibinfo {author} {\bibfnamefont {S.~C.}\ \bibnamefont
  {Zhang}},\ }\bibfield  {title} {\bibinfo {title} {Dynamic spin fluctuations
  and the bag mechanism of high-\$\{\vphantom\}{{T}}\vphantom\{\}\_\{c\}\$
  superconductivity},\ }\href {https://doi.org/10.1103/PhysRevB.39.11663}
  {\bibfield  {journal} {\bibinfo  {journal} {Phys. Rev. B}\ }\textbf {\bibinfo
  {volume} {39}},\ \bibinfo {pages} {11663} (\bibinfo {year}
  {1989})}\BibitemShut {NoStop}%
\bibitem [{\citenamefont {Schrieffer}(1995)}]{Schrieffer1995}%
  \BibitemOpen
  \bibfield  {author} {\bibinfo {author} {\bibfnamefont {J.~R.}\ \bibnamefont
  {Schrieffer}},\ }\bibfield  {title} {\bibinfo {title} {Ward's identity and
  the suppression of spin fluctuation superconductivity},\ }\href
  {https://doi.org/10.1007/BF00752315} {\bibfield  {journal} {\bibinfo
  {journal} {J. Low Temp. Phys.}\ }\textbf {\bibinfo {volume} {99}},\ \bibinfo
  {pages} {397} (\bibinfo {year} {1995})}\BibitemShut {NoStop}%
\bibitem [{\citenamefont {Sachdev}\ \emph {et~al.}(1995)\citenamefont
  {Sachdev}, \citenamefont {Chubukov},\ and\ \citenamefont
  {Sokol}}]{Sokol1995}%
  \BibitemOpen
  \bibfield  {author} {\bibinfo {author} {\bibfnamefont {S.}~\bibnamefont
  {Sachdev}}, \bibinfo {author} {\bibfnamefont {A.~V.}\ \bibnamefont
  {Chubukov}},\ and\ \bibinfo {author} {\bibfnamefont {A.}~\bibnamefont
  {Sokol}},\ }\bibfield  {title} {\bibinfo {title} {Crossover and scaling in a
  nearly antiferromagnetic {{Fermi}} liquid in two dimensions},\ }\href
  {https://doi.org/10.1103/PhysRevB.51.14874} {\bibfield  {journal} {\bibinfo
  {journal} {Phys. Rev. B}\ }\textbf {\bibinfo {volume} {51}},\ \bibinfo
  {pages} {14874} (\bibinfo {year} {1995})}\BibitemShut {NoStop}%
\bibitem [{\citenamefont {Chubukov}\ \emph {et~al.}(1997)\citenamefont
  {Chubukov}, \citenamefont {Monthoux},\ and\ \citenamefont {Morr}}]{Morr1997}%
  \BibitemOpen
  \bibfield  {author} {\bibinfo {author} {\bibfnamefont {A.~V.}\ \bibnamefont
  {Chubukov}}, \bibinfo {author} {\bibfnamefont {P.}~\bibnamefont {Monthoux}},\
  and\ \bibinfo {author} {\bibfnamefont {D.~K.}\ \bibnamefont {Morr}},\
  }\bibfield  {title} {\bibinfo {title} {Vertex corrections in
  antiferromagnetic spin-fluctuation theories},\ }\href
  {https://doi.org/10.1103/PhysRevB.56.7789} {\bibfield  {journal} {\bibinfo
  {journal} {Phys. Rev. B}\ }\textbf {\bibinfo {volume} {56}},\ \bibinfo
  {pages} {7789} (\bibinfo {year} {1997})}\BibitemShut {NoStop}%
\bibitem [{\citenamefont {Chubukov}\ and\ \citenamefont
  {Morr}(1997)}]{Morr_1997_a}%
  \BibitemOpen
  \bibfield  {author} {\bibinfo {author} {\bibfnamefont {A.~V.}\ \bibnamefont
  {Chubukov}}\ and\ \bibinfo {author} {\bibfnamefont {D.~K.}\ \bibnamefont
  {Morr}},\ }\bibfield  {title} {\bibinfo {title} {Electronic structure of
  underdoped cuprates},\ }\href {https://doi.org/10.1016/S0370-1573(97)00033-1}
  {\bibfield  {journal} {\bibinfo  {journal} {Phys. Rep.}\ }\textbf {\bibinfo
  {volume} {288}},\ \bibinfo {pages} {355} (\bibinfo {year}
  {1997})}\BibitemShut {NoStop}%
\bibitem [{\citenamefont {Flambaum}\ \emph {et~al.}(1994)\citenamefont
  {Flambaum}, \citenamefont {Kuchiev},\ and\ \citenamefont
  {Sushkov}}]{Sushkov}%
  \BibitemOpen
  \bibfield  {author} {\bibinfo {author} {\bibfnamefont {V.}~\bibnamefont
  {Flambaum}}, \bibinfo {author} {\bibfnamefont {M.}~\bibnamefont {Kuchiev}},\
  and\ \bibinfo {author} {\bibfnamefont {O.}~\bibnamefont {Sushkov}},\
  }\bibfield  {title} {\bibinfo {title} {Hole-hole superconducting pairing in
  the t-{{J}} model induced by long-range spin-wave exchange},\ }\href
  {https://doi.org/10.1016/0921-4534(94)90081-7} {\bibfield  {journal}
  {\bibinfo  {journal} {Phys. C Supercond.}\ }\textbf {\bibinfo {volume}
  {227}},\ \bibinfo {pages} {267} (\bibinfo {year} {1994})}\BibitemShut
  {NoStop}%
\bibitem [{\citenamefont {Kuchiev}\ and\ \citenamefont
  {Sushkov}(1993)}]{Sushkov_1}%
  \BibitemOpen
  \bibfield  {author} {\bibinfo {author} {\bibfnamefont {{\relax
  M.Yu}.}~\bibnamefont {Kuchiev}}\ and\ \bibinfo {author} {\bibfnamefont
  {O.}~\bibnamefont {Sushkov}},\ }\bibfield  {title} {\bibinfo {title}
  {Large-size two-hole bound states in the t-{{J}} model},\ }\href
  {https://doi.org/10.1016/0921-4534(93)90283-V} {\bibfield  {journal}
  {\bibinfo  {journal} {Phys. C Supercond.}\ }\textbf {\bibinfo {volume}
  {218}},\ \bibinfo {pages} {197} (\bibinfo {year} {1993})}\BibitemShut
  {NoStop}%
\bibitem [{\citenamefont {Ismer}\ \emph {et~al.}(2010)\citenamefont {Ismer},
  \citenamefont {Eremin}, \citenamefont {Rossi}, \citenamefont {Morr},\ and\
  \citenamefont {Blumberg}}]{Ismer_2010}%
  \BibitemOpen
  \bibfield  {author} {\bibinfo {author} {\bibfnamefont {J.-P.}\ \bibnamefont
  {Ismer}}, \bibinfo {author} {\bibfnamefont {I.}~\bibnamefont {Eremin}},
  \bibinfo {author} {\bibfnamefont {E.}~\bibnamefont {Rossi}}, \bibinfo
  {author} {\bibfnamefont {D.~K.}\ \bibnamefont {Morr}},\ and\ \bibinfo
  {author} {\bibfnamefont {G.}~\bibnamefont {Blumberg}},\ }\bibfield  {title}
  {\bibinfo {title} {Theory of multiband superconductivity in spin-density-wave
  metals},\ }\href {https://doi.org/10.1103/PhysRevLett.105.037003} {\bibfield
  {journal} {\bibinfo  {journal} {Phys. Rev. Lett.}\ }\textbf {\bibinfo
  {volume} {105}},\ \bibinfo {pages} {037003} (\bibinfo {year}
  {2010})}\BibitemShut {NoStop}%
\bibitem [{\citenamefont {Adler}(1965)}]{Adler1965}%
  \BibitemOpen
  \bibfield  {author} {\bibinfo {author} {\bibfnamefont {S.~L.}\ \bibnamefont
  {Adler}},\ }\bibfield  {title} {\bibinfo {title} {Consistency {{Conditions}}
  on the {{Strong Interactions Implied}} by a {{Partially Conserved
  Axial-Vector Current}}. {{II}}},\ }\href
  {https://doi.org/10.1103/PhysRev.139.B1638} {\bibfield  {journal} {\bibinfo
  {journal} {Phys. Rev.}\ }\textbf {\bibinfo {volume} {139}},\ \bibinfo {pages}
  {B1638} (\bibinfo {year} {1965})}\BibitemShut {NoStop}%
\bibitem [{\citenamefont {Watanabe}\ and\ \citenamefont
  {Vishwanath}(2014)}]{Watanabe2014}%
  \BibitemOpen
  \bibfield  {author} {\bibinfo {author} {\bibfnamefont {H.}~\bibnamefont
  {Watanabe}}\ and\ \bibinfo {author} {\bibfnamefont {A.}~\bibnamefont
  {Vishwanath}},\ }\bibfield  {title} {\bibinfo {title} {Criterion for
  stability of {{Goldstone}} modes and {{Fermi}} liquid behavior in a metal
  with broken symmetry},\ }\href {https://doi.org/10.1073/pnas.1415592111}
  {\bibfield  {journal} {\bibinfo  {journal} {Proc. Natl. Acad. Sci.}\ }\textbf
  {\bibinfo {volume} {111}},\ \bibinfo {pages} {16314} (\bibinfo {year}
  {2014})}\BibitemShut {NoStop}%
\bibitem [{\citenamefont {Vasiliou}\ \emph {et~al.}(2024)\citenamefont
  {Vasiliou}, \citenamefont {He},\ and\ \citenamefont
  {Bultinck}}]{Vasiliou2024}%
  \BibitemOpen
  \bibfield  {author} {\bibinfo {author} {\bibfnamefont {K.}~\bibnamefont
  {Vasiliou}}, \bibinfo {author} {\bibfnamefont {Y.}~\bibnamefont {He}},\ and\
  \bibinfo {author} {\bibfnamefont {N.}~\bibnamefont {Bultinck}},\ }\bibfield
  {title} {\bibinfo {title} {Electrons interacting with {{Goldstone}} modes and
  the rotating frame},\ }\href {https://doi.org/10.1103/PhysRevB.109.045155}
  {\bibfield  {journal} {\bibinfo  {journal} {Phys. Rev. B}\ }\textbf {\bibinfo
  {volume} {109}},\ \bibinfo {pages} {045155} (\bibinfo {year}
  {2024})}\BibitemShut {NoStop}%
\bibitem [{Note1()}]{Note1}%
  \BibitemOpen
  \bibinfo {note} {The same holds for the pairing by an acoustic phonon. The
  vanishing of the vertex function at $q=0$ implies that the pairing by an
  acoustic phonon is not advantageous compared to the one by an optical
  phonon.}\BibitemShut {Stop}%
\bibitem [{\citenamefont {Kozii}\ \emph {et~al.}(2022)\citenamefont {Kozii},
  \citenamefont {Zaletel},\ and\ \citenamefont {Bultinck}}]{Kozii2022}%
  \BibitemOpen
  \bibfield  {author} {\bibinfo {author} {\bibfnamefont {V.}~\bibnamefont
  {Kozii}}, \bibinfo {author} {\bibfnamefont {M.~P.}\ \bibnamefont {Zaletel}},\
  and\ \bibinfo {author} {\bibfnamefont {N.}~\bibnamefont {Bultinck}},\
  }\bibfield  {title} {\bibinfo {title} {Spin-triplet superconductivity from
  intervalley {{Goldstone}} modes in magic-angle graphene},\ }\href
  {https://doi.org/10.1103/PhysRevB.106.235157} {\bibfield  {journal} {\bibinfo
   {journal} {Phys. Rev. B}\ }\textbf {\bibinfo {volume} {106}},\ \bibinfo
  {pages} {235157} (\bibinfo {year} {2022})}\BibitemShut {NoStop}%
\bibitem [{Note2()}]{Note2}%
  \BibitemOpen
  \bibinfo {note} {In TBG, the low-energy fermionic excitations are nearly
  valley-degenerate, hence inter-valley scattering by a Golstone boson in an
  IVC-ordered state is again a Fermi surface process}\BibitemShut {NoStop}%
\bibitem [{\citenamefont {Fay}\ and\ \citenamefont {Appel}(1980)}]{Fay1980}%
  \BibitemOpen
  \bibfield  {author} {\bibinfo {author} {\bibfnamefont {D.}~\bibnamefont
  {Fay}}\ and\ \bibinfo {author} {\bibfnamefont {J.}~\bibnamefont {Appel}},\
  }\bibfield  {title} {\bibinfo {title} {Coexistence of p -state
  superconductivity and itinerant ferromagnetism},\ }\href
  {https://doi.org/10.1103/PhysRevB.22.3173} {\bibfield  {journal} {\bibinfo
  {journal} {Phys. Rev. B}\ }\textbf {\bibinfo {volume} {22}},\ \bibinfo
  {pages} {3173} (\bibinfo {year} {1980})}\BibitemShut {NoStop}%
\bibitem [{\citenamefont {Kirkpatrick}\ \emph {et~al.}(2001)\citenamefont
  {Kirkpatrick}, \citenamefont {Belitz}, \citenamefont {Vojta},\ and\
  \citenamefont {Narayanan}}]{Kirkpatrick2001}%
  \BibitemOpen
  \bibfield  {author} {\bibinfo {author} {\bibfnamefont {T.~R.}\ \bibnamefont
  {Kirkpatrick}}, \bibinfo {author} {\bibfnamefont {D.}~\bibnamefont {Belitz}},
  \bibinfo {author} {\bibfnamefont {T.}~\bibnamefont {Vojta}},\ and\ \bibinfo
  {author} {\bibfnamefont {R.}~\bibnamefont {Narayanan}},\ }\bibfield  {title}
  {\bibinfo {title} {Strong {{Enhancement}} of {{Superconducting T}} c in
  {{Ferromagnetic Phases}}},\ }\href
  {https://doi.org/10.1103/PhysRevLett.87.127003} {\bibfield  {journal}
  {\bibinfo  {journal} {Phys. Rev. Lett.}\ }\textbf {\bibinfo {volume} {87}},\
  \bibinfo {pages} {127003} (\bibinfo {year} {2001})}\BibitemShut {NoStop}%
\bibitem [{\citenamefont {Guo}\ \emph {et~al.}(2024)\citenamefont {Guo},
  \citenamefont {Pack}, \citenamefont {Swann}, \citenamefont {Holtzman},
  \citenamefont {Cothrine}, \citenamefont {Watanabe}, \citenamefont
  {Taniguchi}, \citenamefont {Mandrus}, \citenamefont {Barmak}, \citenamefont
  {Hone}, \citenamefont {Millis}, \citenamefont {Pasupathy},\ and\
  \citenamefont {Dean}}]{Pasupathy2024}%
  \BibitemOpen
  \bibfield  {author} {\bibinfo {author} {\bibfnamefont {Y.}~\bibnamefont
  {Guo}}, \bibinfo {author} {\bibfnamefont {J.}~\bibnamefont {Pack}}, \bibinfo
  {author} {\bibfnamefont {J.}~\bibnamefont {Swann}}, \bibinfo {author}
  {\bibfnamefont {L.}~\bibnamefont {Holtzman}}, \bibinfo {author}
  {\bibfnamefont {M.}~\bibnamefont {Cothrine}}, \bibinfo {author}
  {\bibfnamefont {K.}~\bibnamefont {Watanabe}}, \bibinfo {author}
  {\bibfnamefont {T.}~\bibnamefont {Taniguchi}}, \bibinfo {author}
  {\bibfnamefont {D.}~\bibnamefont {Mandrus}}, \bibinfo {author} {\bibfnamefont
  {K.}~\bibnamefont {Barmak}}, \bibinfo {author} {\bibfnamefont
  {J.}~\bibnamefont {Hone}}, \bibinfo {author} {\bibfnamefont {A.~J.}\
  \bibnamefont {Millis}}, \bibinfo {author} {\bibfnamefont {A.~N.}\
  \bibnamefont {Pasupathy}},\ and\ \bibinfo {author} {\bibfnamefont {C.~R.}\
  \bibnamefont {Dean}},\ }\href@noop {} {\bibinfo {title} {Superconductivity in
  twisted bilayer {{WSe}}{$_{2}$}}} (\bibinfo {year} {2024}),\ \Eprint
  {https://arxiv.org/abs/2406.03418} {arXiv:2406.03418 [cond-mat.mes-hall]}
  \BibitemShut {NoStop}%
\bibitem [{\citenamefont {Xia}\ \emph {et~al.}(2024)\citenamefont {Xia},
  \citenamefont {Han}, \citenamefont {Watanabe}, \citenamefont {Taniguchi},
  \citenamefont {Shan},\ and\ \citenamefont {Mak}}]{Mak2024}%
  \BibitemOpen
  \bibfield  {author} {\bibinfo {author} {\bibfnamefont {Y.}~\bibnamefont
  {Xia}}, \bibinfo {author} {\bibfnamefont {Z.}~\bibnamefont {Han}}, \bibinfo
  {author} {\bibfnamefont {K.}~\bibnamefont {Watanabe}}, \bibinfo {author}
  {\bibfnamefont {T.}~\bibnamefont {Taniguchi}}, \bibinfo {author}
  {\bibfnamefont {J.}~\bibnamefont {Shan}},\ and\ \bibinfo {author}
  {\bibfnamefont {K.~F.}\ \bibnamefont {Mak}},\ }\bibfield  {title} {\bibinfo
  {title} {Superconductivity in twisted bilayer {{WSe2}}},\ }\bibfield
  {journal} {\bibinfo  {journal} {Nature}\ }\href
  {https://doi.org/10.1038/s41586-024-08116-2} {10.1038/s41586-024-08116-2}
  (\bibinfo {year} {2024})\BibitemShut {NoStop}%
\bibitem [{\citenamefont {Dong}\ \emph {et~al.}(2024)\citenamefont {Dong},
  \citenamefont {{Lantagne-Hurtubise}},\ and\ \citenamefont
  {Alicea}}]{Dong2024}%
  \BibitemOpen
  \bibfield  {author} {\bibinfo {author} {\bibfnamefont {Z.}~\bibnamefont
  {Dong}}, \bibinfo {author} {\bibfnamefont {{\'E}.}~\bibnamefont
  {{Lantagne-Hurtubise}}},\ and\ \bibinfo {author} {\bibfnamefont
  {J.}~\bibnamefont {Alicea}},\ }\href@noop {} {\bibinfo {title}
  {Superconductivity from spin-canting fluctuations in rhombohedral graphene}}
  (\bibinfo {year} {2024}),\ \Eprint {https://arxiv.org/abs/2406.17036}
  {arXiv:2406.17036 [cond-mat]} \BibitemShut {NoStop}%
\bibitem [{\citenamefont {Fernandes}\ and\ \citenamefont
  {Chubukov}(2016)}]{Fernandes2017}%
  \BibitemOpen
  \bibfield  {author} {\bibinfo {author} {\bibfnamefont {R.~M.}\ \bibnamefont
  {Fernandes}}\ and\ \bibinfo {author} {\bibfnamefont {A.~V.}\ \bibnamefont
  {Chubukov}},\ }\bibfield  {title} {\bibinfo {title} {Low-energy microscopic
  models for iron-based superconductors: A review},\ }\href
  {https://doi.org/10.1088/1361-6633/80/1/014503} {\bibfield  {journal}
  {\bibinfo  {journal} {Rep. Prog. Phys.}\ }\textbf {\bibinfo {volume} {80}},\
  \bibinfo {pages} {014503} (\bibinfo {year} {2016})}\BibitemShut {NoStop}%
\bibitem [{\citenamefont {Fernandes}\ \emph {et~al.}(2022)\citenamefont
  {Fernandes}, \citenamefont {Coldea}, \citenamefont {Ding}, \citenamefont
  {Fisher}, \citenamefont {Hirschfeld},\ and\ \citenamefont
  {Kotliar}}]{Fernandes2022}%
  \BibitemOpen
  \bibfield  {author} {\bibinfo {author} {\bibfnamefont {R.~M.}\ \bibnamefont
  {Fernandes}}, \bibinfo {author} {\bibfnamefont {A.~I.}\ \bibnamefont
  {Coldea}}, \bibinfo {author} {\bibfnamefont {H.}~\bibnamefont {Ding}},
  \bibinfo {author} {\bibfnamefont {I.~R.}\ \bibnamefont {Fisher}}, \bibinfo
  {author} {\bibfnamefont {P.~J.}\ \bibnamefont {Hirschfeld}},\ and\ \bibinfo
  {author} {\bibfnamefont {G.}~\bibnamefont {Kotliar}},\ }\bibfield  {title}
  {\bibinfo {title} {Iron pnictides and chalcogenides: A new paradigm for
  superconductivity},\ }\href {https://doi.org/10.1038/s41586-021-04073-2}
  {\bibfield  {journal} {\bibinfo  {journal} {Nature}\ }\textbf {\bibinfo
  {volume} {601}},\ \bibinfo {pages} {35} (\bibinfo {year} {2022})}\BibitemShut
  {NoStop}%
\bibitem [{Note3()}]{Note3}%
  \BibitemOpen
  \bibinfo {note} {One can easily make sure that processes involving more than
  two magnons give a smaller contribution to the pairing interaction and can
  therefore be neglected.}\BibitemShut {Stop}%
\bibitem [{Note4()}]{Note4}%
  \BibitemOpen
  \bibinfo {note} {A similar scenario has been proposed for superconductivity
  induced by either SOC or a magnetic field in the non-magnetic phase near the
  onset of an order~\cite {Dong2023a}.}\BibitemShut {Stop}%
\bibitem [{Note5()}]{Note5}%
  \BibitemOpen
  \bibinfo {note} {Another possibility is a spin-density-wave order with
  momentum $K-K'$ (a spin inter-valley coherence order). For $U_1 + U_3 > U_2$,
  which we assume to hold, a homogeneous FM order is the leading
  instability.}\BibitemShut {Stop}%
\bibitem [{\citenamefont {Raines}\ \emph
  {et~al.}(2024{\natexlab{a}})\citenamefont {Raines}, \citenamefont {Glazman},\
  and\ \citenamefont {Chubukov}}]{Raines2024a}%
  \BibitemOpen
  \bibfield  {author} {\bibinfo {author} {\bibfnamefont {Z.~M.}\ \bibnamefont
  {Raines}}, \bibinfo {author} {\bibfnamefont {L.~I.}\ \bibnamefont
  {Glazman}},\ and\ \bibinfo {author} {\bibfnamefont {A.~V.}\ \bibnamefont
  {Chubukov}},\ }\bibfield  {title} {\bibinfo {title} {Unconventional
  discontinuous transitions in a two-dimensional system with spin and valley
  degrees of freedom},\ }\href {https://doi.org/10.1103/PhysRevB.110.155402}
  {\bibfield  {journal} {\bibinfo  {journal} {Phys. Rev. B}\ }\textbf {\bibinfo
  {volume} {110}},\ \bibinfo {pages} {155402} (\bibinfo {year}
  {2024}{\natexlab{a}})},\ \Eprint {https://arxiv.org/abs/2406.04416}
  {arXiv:2406.04416 [cond-mat]} \BibitemShut {NoStop}%
\bibitem [{\citenamefont {Raines}\ \emph
  {et~al.}(2024{\natexlab{b}})\citenamefont {Raines}, \citenamefont {Glazman},\
  and\ \citenamefont {Chubukov}}]{Raines2024b}%
  \BibitemOpen
  \bibfield  {author} {\bibinfo {author} {\bibfnamefont {Z.~M.}\ \bibnamefont
  {Raines}}, \bibinfo {author} {\bibfnamefont {L.~I.}\ \bibnamefont
  {Glazman}},\ and\ \bibinfo {author} {\bibfnamefont {A.~V.}\ \bibnamefont
  {Chubukov}},\ }\bibfield  {title} {\bibinfo {title} {Unconventional
  {{Discontinuous Transitions}} in {{Isospin Systems}}},\ }\href
  {https://doi.org/10.1103/PhysRevLett.133.146501} {\bibfield  {journal}
  {\bibinfo  {journal} {Phys. Rev. Lett.}\ }\textbf {\bibinfo {volume} {133}},\
  \bibinfo {pages} {146501} (\bibinfo {year} {2024}{\natexlab{b}})},\ \Eprint
  {https://arxiv.org/abs/2406.04415} {arXiv:2406.04415 [cond-mat]} \BibitemShut
  {NoStop}%
\bibitem [{\citenamefont {Raines}\ and\ \citenamefont
  {Chubukov}(2024)}]{Raines2024c}%
  \BibitemOpen
  \bibfield  {author} {\bibinfo {author} {\bibfnamefont {Z.~M.}\ \bibnamefont
  {Raines}}\ and\ \bibinfo {author} {\bibfnamefont {A.~V.}\ \bibnamefont
  {Chubukov}},\ }\bibfield  {title} {\bibinfo {title} {Two-dimensional
  {{Stoner}} transitions beyond mean field},\ }\href
  {https://doi.org/10.1103/PhysRevB.110.235433} {\bibfield  {journal} {\bibinfo
   {journal} {Phys. Rev. B}\ }\textbf {\bibinfo {volume} {110}},\ \bibinfo
  {pages} {235433} (\bibinfo {year} {2024})},\ \Eprint
  {https://arxiv.org/abs/2409.18934} {arXiv:2409.18934 [cond-mat.str-el]}
  \BibitemShut {NoStop}%
\bibitem [{\citenamefont {Calvera}\ \emph {et~al.}(2024)\citenamefont
  {Calvera}, \citenamefont {Valenti}, \citenamefont {Huber}, \citenamefont
  {Berg},\ and\ \citenamefont {Kivelson}}]{Calvera2024}%
  \BibitemOpen
  \bibfield  {author} {\bibinfo {author} {\bibfnamefont {V.}~\bibnamefont
  {Calvera}}, \bibinfo {author} {\bibfnamefont {A.}~\bibnamefont {Valenti}},
  \bibinfo {author} {\bibfnamefont {S.~D.}\ \bibnamefont {Huber}}, \bibinfo
  {author} {\bibfnamefont {E.}~\bibnamefont {Berg}},\ and\ \bibinfo {author}
  {\bibfnamefont {S.~A.}\ \bibnamefont {Kivelson}},\ }\href
  {https://doi.org/10.48550/arXiv.2406.12825} {\bibinfo {title} {Theory of
  {{Coulomb}} driven nematicity in a multi-valley two-dimensional electron
  gas}} (\bibinfo {year} {2024}),\ \Eprint {https://arxiv.org/abs/2406.12825}
  {arXiv:2406.12825 [cond-mat]} \BibitemShut {NoStop}%
\bibitem [{\citenamefont {Dong}\ \emph
  {et~al.}(2023{\natexlab{a}})\citenamefont {Dong}, \citenamefont {Levitov},\
  and\ \citenamefont {Chubukov}}]{Dong2023b}%
  \BibitemOpen
  \bibfield  {author} {\bibinfo {author} {\bibfnamefont {Z.}~\bibnamefont
  {Dong}}, \bibinfo {author} {\bibfnamefont {L.}~\bibnamefont {Levitov}},\ and\
  \bibinfo {author} {\bibfnamefont {A.~V.}\ \bibnamefont {Chubukov}},\
  }\bibfield  {title} {\bibinfo {title} {Superconductivity near spin and valley
  orders in graphene multilayers},\ }\href
  {https://doi.org/10.1103/PhysRevB.108.134503} {\bibfield  {journal} {\bibinfo
   {journal} {Phys. Rev. B}\ }\textbf {\bibinfo {volume} {108}},\ \bibinfo
  {pages} {134503} (\bibinfo {year} {2023}{\natexlab{a}})}\BibitemShut
  {NoStop}%
\bibitem [{Note6()}]{Note6}%
  \BibitemOpen
  \bibinfo {note} {We are thankful to Erez Berg for emphasizing that $A(0,0)$
  must vanish.}\BibitemShut {Stop}%
\bibitem [{\citenamefont {Coleman}(2015)}]{Coleman2015}%
  \BibitemOpen
  \bibfield  {author} {\bibinfo {author} {\bibfnamefont {P.}~\bibnamefont
  {Coleman}},\ }\href {https://doi.org/10.1017/cbo9781139020916} {\emph
  {\bibinfo {title} {Introduction to Many-Body Physics}}}\ (\bibinfo
  {publisher} {Cambridge University Press},\ \bibinfo {year}
  {2015})\BibitemShut {NoStop}%
\bibitem [{Note7()}]{Note7}%
  \BibitemOpen
  \bibinfo {note} {The authors of [\protect \rev@citealp {Dong2024}] also
  speculated that although the theoretical $\beta $ contains a small ratio
  ($U_3/U_1$ in our case), it may be $O(1)$ for the actual parameters. In this
  situation, their $\delta k \sim k_F$, as is generally expected, and also our
  $\lambda ^{sc}_{2mag}$ is not reduced by $\beta $. However, for $\beta =
  O(1)$, there is no clear distinction between ``low-energy'' and
  ``high-energy'' contributions and a repulsive $U_2$ becomes more detrimental
  for superconductivity.}\BibitemShut {Stop}%
\bibitem [{\citenamefont {Chou}\ \emph {et~al.}(2025)\citenamefont {Chou},
  \citenamefont {Zhu},\ and\ \citenamefont {Das~Sarma}}]{Chou2025}%
  \BibitemOpen
  \bibfield  {author} {\bibinfo {author} {\bibfnamefont {Y.-Z.}\ \bibnamefont
  {Chou}}, \bibinfo {author} {\bibfnamefont {J.}~\bibnamefont {Zhu}},\ and\
  \bibinfo {author} {\bibfnamefont {S.}~\bibnamefont {Das~Sarma}},\ }\bibfield
  {title} {\bibinfo {title} {Intravalley spin-polarized superconductivity in
  rhombohedral tetralayer graphene},\ }\href
  {https://doi.org/10.1103/PhysRevB.111.174523} {\bibfield  {journal} {\bibinfo
   {journal} {Phys. Rev. B}\ }\textbf {\bibinfo {volume} {111}},\ \bibinfo
  {pages} {174523} (\bibinfo {year} {2025})}\BibitemShut {NoStop}%
\bibitem [{\citenamefont {Geier}\ \emph {et~al.}(2024)\citenamefont {Geier},
  \citenamefont {Davydova},\ and\ \citenamefont {Fu}}]{Geier2024}%
  \BibitemOpen
  \bibfield  {author} {\bibinfo {author} {\bibfnamefont {M.}~\bibnamefont
  {Geier}}, \bibinfo {author} {\bibfnamefont {M.}~\bibnamefont {Davydova}},\
  and\ \bibinfo {author} {\bibfnamefont {L.}~\bibnamefont {Fu}},\ }\href@noop
  {} {\bibinfo {title} {Chiral and topological superconductivity in isospin
  polarized multilayer graphene}} (\bibinfo {year} {2024}),\ \Eprint
  {https://arxiv.org/abs/2409.13829} {arXiv:2409.13829 [cond-mat.supr-con]}
  \BibitemShut {NoStop}%
\bibitem [{\citenamefont {Yang}\ and\ \citenamefont {Zhang}(2024)}]{Yang2024}%
  \BibitemOpen
  \bibfield  {author} {\bibinfo {author} {\bibfnamefont {H.}~\bibnamefont
  {Yang}}\ and\ \bibinfo {author} {\bibfnamefont {Y.-H.}\ \bibnamefont
  {Zhang}},\ }\href@noop {} {\bibinfo {title} {Topological incommensurate
  {{Fulde-Ferrell-Larkin-Ovchinnikov}} superconductor and {{Bogoliubov Fermi}}
  surface in rhombohedral tetra-layer graphene}} (\bibinfo {year} {2024}),\
  \Eprint {https://arxiv.org/abs/2411.02503} {arXiv:2411.02503
  [cond-mat.supr-con]} \BibitemShut {NoStop}%
\bibitem [{\citenamefont {Qin}\ and\ \citenamefont {Wu}(2024)}]{Qin2024}%
  \BibitemOpen
  \bibfield  {author} {\bibinfo {author} {\bibfnamefont {Q.}~\bibnamefont
  {Qin}}\ and\ \bibinfo {author} {\bibfnamefont {C.}~\bibnamefont {Wu}},\
  }\href@noop {} {\bibinfo {title} {Chiral finite-momentum superconductivity in
  the tetralayer graphene}} (\bibinfo {year} {2024}),\ \Eprint
  {https://arxiv.org/abs/2412.07145} {arXiv:2412.07145 [cond-mat.supr-con]}
  \BibitemShut {NoStop}%
\bibitem [{\citenamefont {Jahin}\ and\ \citenamefont {Lin}(2025)}]{Jahin2025}%
  \BibitemOpen
  \bibfield  {author} {\bibinfo {author} {\bibfnamefont {A.}~\bibnamefont
  {Jahin}}\ and\ \bibinfo {author} {\bibfnamefont {S.-Z.}\ \bibnamefont
  {Lin}},\ }\href@noop {} {\bibinfo {title} {Enhanced {{Kohn-Luttinger}}
  topological superconductivity in bands with nontrivial geometry}} (\bibinfo
  {year} {2025}),\ \Eprint {https://arxiv.org/abs/2411.09664} {arXiv:2411.09664
  [cond-mat.supr-con]} \BibitemShut {NoStop}%
\bibitem [{\citenamefont {Gil}\ and\ \citenamefont {Berg}(2025)}]{Gil2025}%
  \BibitemOpen
  \bibfield  {author} {\bibinfo {author} {\bibfnamefont {A.}~\bibnamefont
  {Gil}}\ and\ \bibinfo {author} {\bibfnamefont {E.}~\bibnamefont {Berg}},\
  }\href {https://doi.org/10.48550/arXiv.2504.19321} {\bibinfo {title} {Charge
  and pair density waves in a spin and valley-polarized system at a
  {{Van-Hove}} singularity}} (\bibinfo {year} {2025}),\ \Eprint
  {https://arxiv.org/abs/2504.19321} {arXiv:2504.19321 [cond-mat]} \BibitemShut
  {NoStop}%
\bibitem [{\citenamefont {Dong}\ and\ \citenamefont {Lee}(2025)}]{Dong2025}%
  \BibitemOpen
  \bibfield  {author} {\bibinfo {author} {\bibfnamefont {Z.}~\bibnamefont
  {Dong}}\ and\ \bibinfo {author} {\bibfnamefont {P.~A.}\ \bibnamefont {Lee}},\
  }\href {https://doi.org/10.48550/arXiv.2503.11079} {\bibinfo {title} {A
  controllable theory of superconductivity due to strong repulsion in a
  polarized band}} (\bibinfo {year} {2025}),\ \Eprint
  {https://arxiv.org/abs/2503.11079} {arXiv:2503.11079 [cond-mat]} \BibitemShut
  {NoStop}%
\bibitem [{\citenamefont {Gaggioli}\ \emph {et~al.}(2025)\citenamefont
  {Gaggioli}, \citenamefont {Guerci},\ and\ \citenamefont {Fu}}]{Gaggioli2025}%
  \BibitemOpen
  \bibfield  {author} {\bibinfo {author} {\bibfnamefont {F.}~\bibnamefont
  {Gaggioli}}, \bibinfo {author} {\bibfnamefont {D.}~\bibnamefont {Guerci}},\
  and\ \bibinfo {author} {\bibfnamefont {L.}~\bibnamefont {Fu}},\ }\href@noop
  {} {\bibinfo {title} {Spontaneous vortex-antivortex lattice and {{Majorana}}
  fermions in rhombohedral graphene}} (\bibinfo {year} {2025}),\ \Eprint
  {https://arxiv.org/abs/2503.16384} {arXiv:2503.16384 [cond-mat.supr-con]}
  \BibitemShut {NoStop}%
\bibitem [{\citenamefont {{Parra-Martinez}}\ \emph {et~al.}(2025)\citenamefont
  {{Parra-Martinez}}, \citenamefont {{Jimeno-Pozo}}, \citenamefont {Phong},
  \citenamefont {{Sainz-Cruz}}, \citenamefont {Kaplan}, \citenamefont
  {Emanuel}, \citenamefont {Oreg}, \citenamefont {Pantaleon}, \citenamefont
  {{Silva-Guillen}},\ and\ \citenamefont {Guinea}}]{Parramartinez2025}%
  \BibitemOpen
  \bibfield  {author} {\bibinfo {author} {\bibfnamefont {G.}~\bibnamefont
  {{Parra-Martinez}}}, \bibinfo {author} {\bibfnamefont {A.}~\bibnamefont
  {{Jimeno-Pozo}}}, \bibinfo {author} {\bibfnamefont {V.~T.}\ \bibnamefont
  {Phong}}, \bibinfo {author} {\bibfnamefont {H.}~\bibnamefont {{Sainz-Cruz}}},
  \bibinfo {author} {\bibfnamefont {D.}~\bibnamefont {Kaplan}}, \bibinfo
  {author} {\bibfnamefont {P.}~\bibnamefont {Emanuel}}, \bibinfo {author}
  {\bibfnamefont {Y.}~\bibnamefont {Oreg}}, \bibinfo {author} {\bibfnamefont
  {P.~A.}\ \bibnamefont {Pantaleon}}, \bibinfo {author} {\bibfnamefont {J.~A.}\
  \bibnamefont {{Silva-Guillen}}},\ and\ \bibinfo {author} {\bibfnamefont
  {F.}~\bibnamefont {Guinea}},\ }\href@noop {} {\bibinfo {title} {Band
  renormalization, quarter metals, and chiral superconductivity in rhombohedral
  tetralayer graphene}} (\bibinfo {year} {2025}),\ \Eprint
  {https://arxiv.org/abs/2502.19474} {arXiv:2502.19474 [cond-mat.str-el]}
  \BibitemShut {NoStop}%
\bibitem [{\citenamefont {Kim}\ \emph {et~al.}(2025)\citenamefont {Kim},
  \citenamefont {Timmel}, \citenamefont {Ju},\ and\ \citenamefont
  {Wen}}]{Kim2025}%
  \BibitemOpen
  \bibfield  {author} {\bibinfo {author} {\bibfnamefont {M.}~\bibnamefont
  {Kim}}, \bibinfo {author} {\bibfnamefont {A.}~\bibnamefont {Timmel}},
  \bibinfo {author} {\bibfnamefont {L.}~\bibnamefont {Ju}},\ and\ \bibinfo
  {author} {\bibfnamefont {X.-G.}\ \bibnamefont {Wen}},\ }\bibfield  {title}
  {\bibinfo {title} {Topological chiral superconductivity beyond pairing in a
  {{Fermi}} liquid},\ }\href {https://doi.org/10.1103/PhysRevB.111.014508}
  {\bibfield  {journal} {\bibinfo  {journal} {Phys. Rev. B}\ }\textbf {\bibinfo
  {volume} {111}},\ \bibinfo {pages} {014508} (\bibinfo {year}
  {2025})}\BibitemShut {NoStop}%
\bibitem [{\citenamefont {Christos}\ \emph {et~al.}(2025)\citenamefont
  {Christos}, \citenamefont {Bonetti},\ and\ \citenamefont
  {Scheurer}}]{Christos2025}%
  \BibitemOpen
  \bibfield  {author} {\bibinfo {author} {\bibfnamefont {M.}~\bibnamefont
  {Christos}}, \bibinfo {author} {\bibfnamefont {P.~M.}\ \bibnamefont
  {Bonetti}},\ and\ \bibinfo {author} {\bibfnamefont {M.~S.}\ \bibnamefont
  {Scheurer}},\ }\href@noop {} {\bibinfo {title} {Finite-momentum pairing and
  superlattice superconductivity in valley-imbalanced rhombohedral graphene}}
  (\bibinfo {year} {2025}),\ \Eprint {https://arxiv.org/abs/2503.15471}
  {arXiv:2503.15471 [cond-mat.str-el]} \BibitemShut {NoStop}%
\bibitem [{\citenamefont {Raines}\ and\ \citenamefont
  {Chubukov}(2025)}]{Raines2025a}%
  \BibitemOpen
  \bibfield  {author} {\bibinfo {author} {\bibfnamefont {Z.}~\bibnamefont
  {Raines}}\ and\ \bibinfo {author} {\bibfnamefont {A.~V.}\ \bibnamefont
  {Chubukov}},\ }\href@noop {} {\bibinfo {title} {To {{Appear}}}} (\bibinfo
  {year} {2025})\BibitemShut {NoStop}%
\bibitem [{\citenamefont {Dong}\ \emph
  {et~al.}(2023{\natexlab{b}})\citenamefont {Dong}, \citenamefont {Chubukov},\
  and\ \citenamefont {Levitov}}]{Dong2023a}%
  \BibitemOpen
  \bibfield  {author} {\bibinfo {author} {\bibfnamefont {Z.}~\bibnamefont
  {Dong}}, \bibinfo {author} {\bibfnamefont {A.~V.}\ \bibnamefont {Chubukov}},\
  and\ \bibinfo {author} {\bibfnamefont {L.}~\bibnamefont {Levitov}},\
  }\bibfield  {title} {\bibinfo {title} {Transformer spin-triplet
  superconductivity at the onset of isospin order in bilayer graphene},\ }\href
  {https://doi.org/10.1103/PhysRevB.107.174512} {\bibfield  {journal} {\bibinfo
   {journal} {Phys. Rev. B}\ }\textbf {\bibinfo {volume} {107}},\ \bibinfo
  {pages} {174512} (\bibinfo {year} {2023}{\natexlab{b}})}\BibitemShut
  {NoStop}%
\end{thebibliography}%

\appendix

\section{Effective interactions mediated by a single magnon}
\label{sec:Ueff}

\begin{figure*}[htb]
    \centering
    \includegraphics[width=0.9\linewidth]{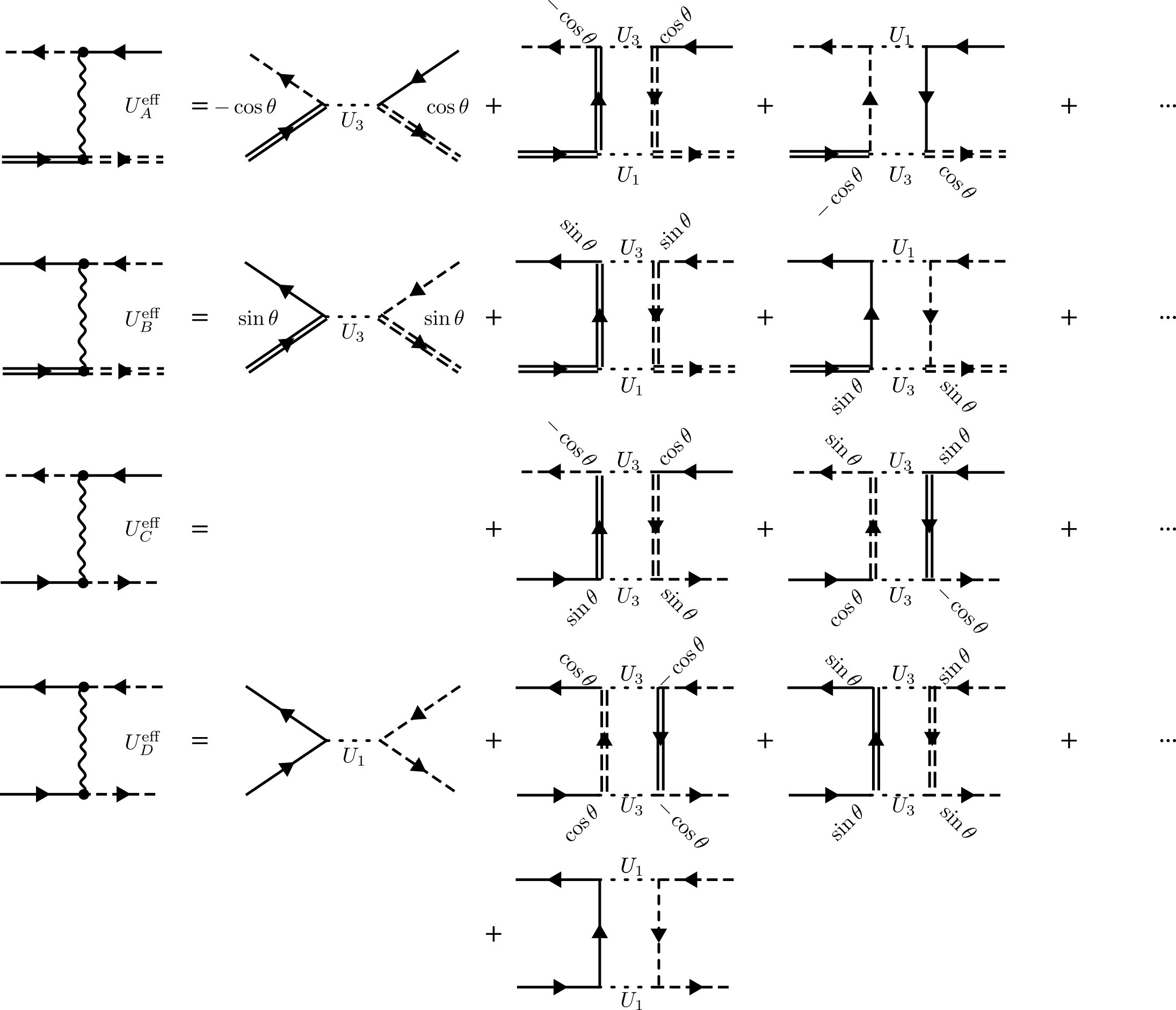}
    \caption{
        Ladder series for the effective interactions $U^\text{eff}_{A-D}$ mediated by a single magnon. We show the
        first few terms in each series.
        The form-factors from
        the rotated $U_3$ interaction are shown
        next to the vertices of the diagrams.}
    \label{fig:ladders}
\end{figure*}

Each of the four interactions $U^\text{eff}_{A-D}$,  mediated by a single magnon, is obtained  by summing up ladder series of diagrams shown in \cref{fig:ladders}.

The full expressions for $U^\text{eff}_{A-D}$, valid for $q<q_c$ and $q > q_c$ are
\begin{equation}
    \begin{split}
        U^\text{eff}_A & = -   U_3 \cos^2{\theta} \frac{(1- U_1 \Pi)^2 - (U_3 \Pi)^2 \cos{2\theta}}{D_+ D_-},                  \\
        U^\text{eff}_B & =     U_3 \sin^2{\theta} \frac{(1- U_1 \Pi)^2 + (U_3 \Pi)^2 \cos{2\theta}}{D_+ D_-},                  \\
        U^\text{eff}_C & =  -  2 U^2_3 \Pi  \sin^2{\theta} \cos^2{\theta}  \frac{(1- U_1\Pi)}{D_+ D_-},                        \\
        U^\text{eff}_D & =     2 U^2_3 \Pi  \sin^2{\theta} \cos^2{\theta}  \frac{(1- U_1\Pi)}{D_+ D_-} + \delta U^\text{eff}_D
    \end{split}
\end{equation}
where
\begin{equation}
    \delta U^\text{eff}_D = U^1_1 \Pi (1- U_1 \Pi) \frac{(1-U_1 \Pi)^2 - (U_3 \Pi)^2}{D_+ D_-}.
\end{equation}
In all terms   we approximated the polarization bubble in the numerator by static uniform
$\Pi = \Pi (0, 0)$.
The interactions $U^\text{eff}_A$ and $U^\text{eff}_C$ are non-zero only in the presence of SOC $\lambda$.  Interaction
$ U^\text{eff}_B$ survives  at $\lambda =0$, but scales as $U_3$.
Interaction $U^\text{eff}_D$ survives when $\lambda$ and $U_3$ are both zero and is the leading term in the analysis of the PDW pairing~\cite{Raines2025a}.
The expression $U^\text{eff}_{B,D} = - U^\text{eff}_{A,C} =  U_3 \cos^2{\theta} \chi_{\perp} (q, \Omega_m)$, which we cited in main text, is obtained by setting $q <q_c$, expanding $D_+$ in $q$ and $\Omega_m$ to quadratic order, and setting $1 -U_1 \Pi = U_3 \Pi$ in the numerator and in $D_-$.

The effective two-fermion/two-magnon vertex is obtained by
summing the diagrams in \cref{fig:two-magnon-scattering}.
If we expand the vertex in the pairing diagrams \cref{fig:two-magnon-diagrams} it can be seen that all four interactions $U^\text{eff}_{A-D}$ contribute.

\section{Effective pairing interaction in terms of the full fermion-two-magnon-vertex}
\label{sec:effective-interaction}

\begin{figure}[htb]
    \centering
    \includegraphics[width=\linewidth]{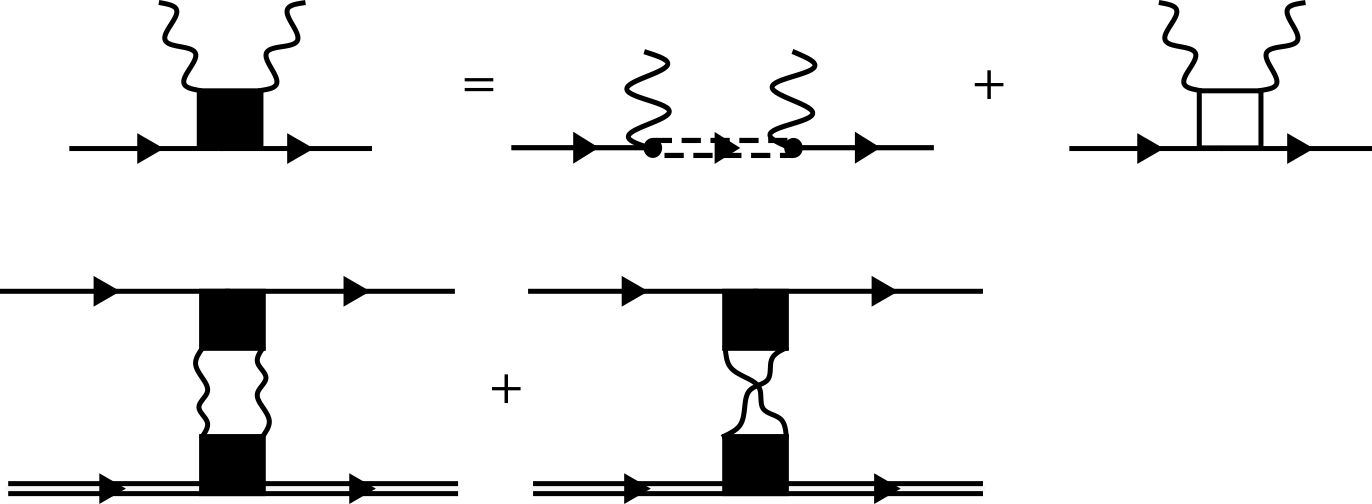}
    \caption{Effective pairing interaction in terms of the full fermion-two-magnon-vertex.}
    \label{fig:effective-interaction}
\end{figure}

An alternative way to think about the effective pairing interactions is to introduce a full effective two-fermion/two-magnon vertex function.
Like the  vertex function above (\cref{fig:two-magnon-diagrams}), the full vertex function is non-local.
We depict the full vertex function and the effective interaction in terms of it in \cref{fig:effective-interaction}.
Note that because the vertex function is non-local, the two diagrams (crossed and uncrossed) are not equivalent.

\end{document}